\documentclass[12pt]{article}
\pdfoutput=1
\usepackage[totalwidth=6.5in,totalheight=9in]{geometry}
\usepackage[centertags]{amsmath}
\usepackage[square,comma,sort&compress,numbers]{natbib}
\usepackage{epsfig}
\usepackage{amsfonts}
\usepackage{amssymb}
\usepackage{slashed}
\usepackage{mathrsfs}
\usepackage{bbm}
\usepackage{array,multirow}
\usepackage{graphicx}
\usepackage{caption}
\usepackage{subcaption}
\usepackage{color}
\usepackage{setspace}
\usepackage{comment}
\usepackage{enumitem}
\usepackage{hyperref}

\newcommand{\ra}{\rightarrow}

\newcommand{\lra}{\leftrightarrow}

\newcommand{\<}{\langle}
\renewcommand{\>}{\rangle}
\newcommand{\be}{\begin{equation}}
\newcommand{\ee}{\end{equation}}
\newcommand{\ba}{\begin{aligned}}
\newcommand{\ea}{\end{aligned}}
\newcommand{\benn}{\begin{equation*}}
\newcommand{\eenn}{\end{equation*}}
\newcommand{\bi}{\begin{itemize}}  
\newcommand{\ei}{\end{itemize}}
\newcommand{\bpm}{\begin{pmatrix}}
\newcommand{\epm}{\end{pmatrix}}

\newcommand{\p}{\partial}

\newcommand{\De}{\Delta}
\newcommand{\de}{\delta}
\newcommand{\ep}{\epsilon}
\newcommand{\vep}{\varepsilon}
\newcommand{\G}{\Gamma}

\newcommand{\s}{\sigma}

\newcommand{\Ccal}{{\mathcal C}}
\newcommand{\Dcal}{{\mathcal D}}
\newcommand{\Ecal}{{\mathcal E}}
\newcommand{\Fcal}{{\mathcal F}}
\newcommand{\Gcal}{{\mathcal G}}
\newcommand{\Hcal}{{\mathcal H}}

\newcommand{\Ncal}{{\mathcal N}}
\newcommand{\Ocal}{{\mathcal O}}

\newcommand{\CFT}{\textrm{CFT} }

\newcommand{\fr}{\frac}
\newcommand{\tfr}{\tfrac}
\newcommand{\half}{\frac{1}{2}}

\newcommand{\comm}[2]{[#1,#2]}

\newcommand{\Chat}{\widehat{C}}
\renewcommand{\;}{\mspace{1mu}}

\numberwithin{equation}{section}

\begin{document}

\begin{titlepage}

\noindent\makebox[\dimexpr\linewidth-1cm\relax][r]{LA-UR-25-31137}

\setcounter{page}{0}

\begin{center}

\vspace*{2cm}

{\bf \Large Energy Correlator Conformal Blocks and Positivity}

\vspace{0.5cm}

{\bf Bianka Me\c{c}aj$^1$, Ian Moult$^2$, Matthew T.\ Walters$^3$, Yuan Xin$^4$}

\vspace{0.5cm}

{\it $^1$ Theoretical Division, Los Alamos National Laboratory, Los Alamos, NM 87545, USA} \\
{\it $^2$ Department of Physics, Yale University, New Haven, CT 06511, USA} \\
{\it $^3$ Maxwell Institute for Mathematical Sciences, Department of Mathematics, \\
Heriot-Watt University, Edinburgh EH14, UK} \\
{\it $^4$ Department of Physics, Carnegie Mellon University, Pittsburgh, PA 15213, USA}\\

\end{center}

\vspace{1cm}

\begin{abstract}

Correlation functions of energy flow operators (energy-energy correlators) are one of the simplest observables in quantum field theory and gravity, with diverse applications ranging from real world collider physics to constraining the space of consistent theories. 
In this paper we further develop the conformal block decomposition of energy-energy correlators in conformal field theories (CFTs), focusing on the source-detector operator product expansion (OPE).
We compute the general conformal blocks in this channel for traceless symmetric operators of arbitrary spin in the background of a scalar source, considering both parity-even and parity-odd contributions.
Motivated by the availability of data from the conformal bootstrap, we analyze the convergence of this source-detector OPE, taking a tensor product of two decoupled CFTs as an elementary example.
Finally, we use positivity of energy correlators to derive novel bounds on OPE coefficients involving the stress-energy tensor in generic CFTs, and demonstrate the application of these bounds in the specific example of the 3d Ising CFT, obtaining new constraints for both parity-even and parity-odd operators.

\end{abstract}

\end{titlepage}

\tableofcontents


\section{Introduction}
\label{sec:Intro}

A natural observable for studying the dynamics of any quantum field theory (QFT), or theory of quantum gravity, is the distribution of energy flux created by a scattering process or excitation of the system \cite{Sterman:1975xv,Basham:1978zq,Basham:1979gh,Basham:1977iq,Basham:1978bw}. On the one hand, such observables provide theoretical abstractions of collider physics observables, making them ideal theoretical playgrounds for improving our understanding of real Standard Model physics at collider experiments. On the other hand, because they are well defined even in theories that do not admit an S-matrix or correlation functions of local operators, they are an ideal observable for studying the space of consistent quantum field theories or theories of quantum gravity. As such, there has been considerable recent interest in these observable from numerous directions. For a recent review, see~\cite{Moult:2025nhu}.

One of the most appealing features of energy correlators in QFT is that they provide a direct connection between observables which can be measured in experiment and correlation functions of local operators \cite{Hofman:2008ar,Belitsky:2013ofa,Belitsky:2013bja,Belitsky:2013xxa}. The energy flow operator can be written in terms of the stress tensor of the theory as
\cite{Sveshnikov:1995vi,Tkachov:1995kk,Korchemsky:1999kt,Hofman:2008ar},
\be
\Ecal(\hat{n}) \equiv \lim_{r\ra\infty} r^{d-2} \int_0^\infty dt \, \hat{n}^i \; T_{0i}(t,r\hat{n}).
\label{eq:EdefNaive}
\ee
Correlation functions of these energy flow operators in a state $|\psi\rangle$
\be
\<\Ecal(\hat{n}_1) \ldots \Ecal(\hat{n}_k)\> \equiv \fr{\<\psi|\Ecal(\hat{n}_1) \cdots \Ecal(\hat{n}_k)|\psi\>}{\<\psi|\psi\>}\,,
\label{eq:ECorrDef}
\ee
can be directly measured from patterns of the energy flux of particles. This provides a sharp relationship between correlation functions of the stress tensor and experimental observables, which has recently been realized \cite{Chen:2020vvp,Komiske:2022enw}. Indeed, the two-point energy correlator in a state created by the electromagnetic current was recently measured to high precision in QCD \cite{Electron-PositronAlliance:2025fhk}. While such correlation functions cannot currently be computed non-perturbatively in QCD, an exciting prospect is that they could be computed non-perturbatively in conformal field theories (CFTs). This would mark the first time that energy flux observables used experimentally can also be non-perturbatively computed in simplified theories. Comparisons between precision measurements of these observables in QCD and non-perturbative calculations in simplified theories can provide valuable insight into non-perturbative phenomena in QCD. 

In the last decade there has been tremendous progress in non-perturbative calculations 
in CFTs, where conformal symmetry provides powerful numerical and analytic tools for the calculation of correlation functions of local operators, see \cite{Poland:2018epd,Hartman:2022zik,Rychkov:2023wsd} for reviews.   The standard approach to studying multi-point correlation functions is the operator product expansion (OPE). For the energy correlator there are two-channels: on the one hand one can study the OPE of two energy flow operators, giving rise to an expansion in detector operators. This is referred to as the light-ray (or ``s-channel'') OPE \cite{Hofman:2008ar}. It has been extensively developed in the context of CFTs \cite{Kologlu:2019mfz,Chang:2020qpj}, and has recently found applications in QCD \cite{Chen:2020adz,Chen:2021gdk,Chen:2022jhb,Chang:2022ryc,Chen:2025rjc}. However, it is sensitive to the spectrum of light-ray operators in the theory. While this is being actively explored, both at weak and finite coupling \cite{Alfimov:2014bwa,Gromov:2015wca,Gromov:2015vua,Caron-Huot:2022eqs,Klabbers:2023zdz,Ekhammar:2024neh,Brizio:2024nso,Li:2025knf,Ekhammar:2025vig}, it is currently under less control than the data for local operators. Alternatively, one can consider the source-detector (or ``t-channel'') OPE \cite{Kologlu:2019bco}, which is an expansion in terms of local operators. Explicitly for the two-point energy correlator in the state produced by a scalar operator, $\phi$, we can write
\be
\Fcal_\Ecal(z)\equiv\fr{\<\Ecal(\hat{n}_1) \; \Ecal(\hat{n}_2)\>}{\<\Ecal(\hat{n}_1)\> \; \<\Ecal(\hat{n}_2)\>} = \sum_\Ocal \lambda_{\phi T\Ocal}^2 \; \Gcal_\Ocal(z),
\label{eq:ConfBlock4d}
\ee
where the sum is over primary operators $\mathcal{O}$. In a scalar state, this observable is a function of the angle between the two detectors, $\hat{n}_1 \cdot \hat{n}_2 \equiv \cos\theta$, which we often characterize using the cross-ratio  $z=\half(1-\cos\theta)$.  The functions $\Gcal_\Ocal(z)$, called ``conformal blocks'', are completely fixed by conformal symmetry and depend only on the scaling dimension and spin quantum numbers of the operator $\Ocal$. They are each multiplied by an associated theory-dependent OPE coefficient $\lambda_{\phi T\Ocal}$ which encodes the dynamics of the CFT. This OPE is appealing in that it directly relates the energy correlator distribution to the spectrum of local operators and OPE coefficients involving the stress tensor.

The numerical conformal bootstrap has been most successful in the calculation of OPE coefficients $\lambda_{\phi_1 \phi_2 \Ocal}$ with two scalars (see e.g.~\cite{Kos:2016ysd,Simmons-Duffin:2016wlq,Chester:2019ifh,Reehorst:2021hmp,Atanasov:2022bpi,Erramilli:2022kgp}), whereas the required OPE coefficients with more than one spinning operator have been challenging to compute. Recently progress has been made in the calculation of such data, using both the numerical bootstrap~\cite{Dymarsky:2017yzx,Reehorst:2019pzi,He:2023ewx,Poland:2023vpn,Poland:2023bny,Chang:2024whx,Poland:2025ide} and fuzzy sphere regularization~\cite{Zhu:2022gjc,Hu:2023xak,Han:2023yyb,Lauchli:2025fii,Fan:2025bhc,ArguelloCruz:2025zuq,EliasMiro:2025msj}. In the specific case of $\mathcal{N}=4$ super Yang-Mills (SYM) there has also been tremendous progress combining the numerical conformal bootstrap with data from the quantum spectral curve \cite{Gromov:2013pga,Gromov:2014caa} or supersymmetric localization \cite{Chester:2020dja,Chester:2021aun,Chester:2025kvw}, yielding impressive results \cite{Cavaglia:2021bnz,Cavaglia:2022qpg,Caron-Huot:2022sdy,Cavaglia:2023mmu,Chester:2023ehi,Caron-Huot:2024tzr,Cavaglia:2024dkk}.

Beautiful examples of using OPE coefficients to reconstruct Euclidean four-point functions include the four-point function of $\sigma$ in the 3d Ising CFT~\cite{Rychkov:2016mrc} and the four-point function of $\phi$ in the $O(2)$ model~\cite{Liu:2020tpf} from the conformal bootstrap, four-point functions of $\sigma$ and $\epsilon$ in the 3d Ising CFT from the fuzzy sphere~\cite{Han:2023yyb}, and the four-point function of $\mathcal{O}_{20'}$ operators in $\mathcal{N}=4$ super Yang-Mills \cite{Caron-Huot:2024tzr}. It is therefore interesting to explore to what extent this data can be used to reconstruct energy correlator observables using the t-channel OPE. Being a Lorentzian observable, it is less clear that this will work directly, and more sophisticated techniques may be required, but we feel that as more OPE data emerges, it is worth exploring and developing its connection with energy flux observables.\footnote{Quite remarkably, in the case of $\mathcal{N}=4$ SYM the energy correlator has been bootstrapped at finite coupling \cite{N4_bootstrap}. We hope that this is the first of many theories in which one can obtain non-perturbative results for energy correlators.}

The goal of this paper is to explore and further develop the t-channel OPE for energy correlator observables.\footnote{The t-channel OPE was previously used in~\cite{Chen:2024iuv} to study energy correlators in $\Ncal=4$ SYM, where supersymmetry relates them to correlators of simpler detectors built from scalar operators.}  In particular, we compute the  conformal blocks appearing in the t-channel OPE for all traceless symmetric operators of arbitrary spin in any spacetime dimension, providing a necessary first step towards the non-perturbative calculation of energy correlators in general CFTs.  Our result extends the calculation of the scalar event shape blocks in \cite{Kologlu:2019bco}, allowing much broader applications, in particular to theories where the stress tensor is not related to a scalar operator by supersymmetry (such as the 3d Ising CFT). We consider both parity-even structures in general dimensions, as well as parity-odd structures specific to $d=3$.  As a simple demonstration of the use of these conformal blocks, we present the example of a tensor product theory $\mathrm{CFT}_1 \otimes \mathrm{CFT}_2$, where the mixed energy-energy correlator factorizes and the necessary OPE coefficients can be computed exactly. This theory also provides a controlled environment in which to examine the convergence of the conformal block expansion. In addition to the energy correlator expressed in terms of the angle between the detectors, as in~\eqref{eq:ConfBlock4d}, we also consider its expansion in partial waves, which in $d=4$ correspond to Legendre polynomials,
\be
\Fcal_\Ecal(z) = \sum_{\ell=0}^\infty \Fcal_\Ecal^{(\ell)} \; (2\ell+1) \; P_\ell(\cos\theta).
\label{eq:PartialWave4d}
\ee
The coefficients in the partial wave expansion are directly related to the OPE coefficients via
\be\label{eq:partialwave_intro}
\Fcal_\Ecal^{(\ell)} = \sum_\Ocal \lambda_{\phi T\Ocal}^2 \; \Gcal_\Ocal^{(\ell)},
\ee
where $\Gcal_\Ocal^{(\ell)} \geq 0$ is the partial wave coefficient of a t-channel conformal block. While less explored in phenomenological studies in collider physics, as compared to the full distribution, these partial waves also characterize the energy distribution in collisions. We find that while the partial wave coefficients for the observable converge well, the convergence for the energy correlator distribution as a function of angle is poor. This suggests an intrinsic challenge in using Euclidean OPE data to reconstruct the energy correlator distribution as a function of angle from local OPE data, and motivates direct Lorentzian bootstrap approaches. Alternatively, it motivates further exploration of the partial wave coefficients in phenomenological studies.

Another interesting feature of~\eqref{eq:ConfBlock4d} is that by relating energy flux observables to the spectrum and OPE coefficients of local operators in the CFT, it enables the use of positivity properties of the energy flow operator to constrain properties of the CFT data. Indeed, this operator famously satisfies the Average Null Energy Condition (ANEC)
\be
\<\psi|\Ecal(\hat{n})|\psi\> \geq 0 \textrm{ for all } |\psi\>\,,
\label{eq:ANEC_intro}
\ee
which has been rigorously proven for general unitary, Lorentz-invariant QFTs~\cite{Faulkner:2016mzt,Hartman:2016lgu}. This was used in \cite{Hofman:2008ar} to derive the ``conformal collider bounds" on anomaly coefficients. These bounds do not arise from crossing, but instead follow simply from energy positivity. However, it was shown that they can also be directly derived from the conformal bootstrap \cite{Hofman:2016awc}. ANEC positivity has also been applied to constrain other CFT data \cite{Li:2015itl,Chowdhury:2017vel,Cordova:2017zej,Cordova:2017dhq}, and most recently, to constrain renormalization group flows \cite{Hartman:2023qdn,Hartman:2023ccw,Hartman:2024xkw,Nakamura:2025hyw}.

While the implications of positivity for the one-point correlator have been extensively explored, higher point positivity of energy correlators
\be
\<\psi|\Ecal(\hat{n}_1) \cdots \Ecal(\hat{n}_k)|\psi\> \geq 0 \textrm{ for all } |\psi\>\,,
\ee
or constraints arising from higher point correlators more generally, have been much less explored. One exception is two beautiful applications for constraining theories with higher spin symmetry \cite{Maldacena:2011jn,Maldacena:2012sf} and CFTs with extremal values of $a/c$ \cite{Zhiboedov:2013opa}. There has recently been significant progress in the calculation of higher-point correlation functions in perturbation theory \cite{Chen:2019bpb,Yan:2022cye,Yang:2022tgm,Chicherin:2024ifn,Yang:2024gcn,Ma:2025qtx}, as well as their measurement \cite{Chen:2022swd}. These advances motivate revisiting how these observables can be used to constrain theories.

Although not directly phrased in the language of higher-point correlators, ref.~\cite{Cordova:2017zej} considered an interesting setup, referred to as ``interference effects" in the conformal collider. They considered the ANEC in states produced by linear combinations of operators
\begin{align}
|\psi\rangle =v_1|\mathcal{O}_1 \rangle +v_2 |\mathcal{O}_2 \rangle\,,
\end{align}
namely
\begin{align}\label{eq:juan_positive}
\langle \psi | \mathcal{E}(\hat{n}) | \psi \rangle = v^\dagger
\begin{pmatrix}
    \langle \mathcal{O}_1 | \mathcal{E}(\hat{n}) | \mathcal{O}_1 \rangle  & \langle \mathcal{O}_1 | \mathcal{E}(\hat{n}) | \mathcal{O}_2 \rangle 
\\
    \langle \mathcal{O}_1 | \mathcal{E}(\hat{n}) | \mathcal{O}_2 \rangle^\dagger  & \langle \mathcal{O}_2 | \mathcal{E}(\hat{n}) | \mathcal{O}_2 \rangle 
  \end{pmatrix}
  v \geq 0\,,
\end{align}
which imposes that the matrix of ANEC expectation values is positive semi-definite.
Ref.~\cite{Cordova:2017zej} specifically considered the stress tensor and a scalar primary operator, however this can be generalized to other operators or to larger linear combinations. The constraints derived using this approach are in general stronger, due to the appearance of the off-diagonal terms. The authors of \cite{Cordova:2017zej} used these constraints to derive a number of interesting constraints on $\mathcal{O}TT$ three-point structures (see also \cite{Meltzer:2017rtf}).

A natural way of phrasing these interference effects is to view them as arising from the t-channel OPE of a higher-point energy correlator. Indeed, this is familiar from the bootstrap for correlation functions of local operators, and provides an immediate way of generalizing~\eqref{eq:juan_positive} to include information from an infinite set of operators. Expressed in terms of operators, the t-channel OPE takes the form of a sum over interference terms of one-point matrix elements of the energy flow operator
\be
\Fcal_\Ecal(z) = \sum_\Ocal \int\fr{d^dp'}{(2\pi)^d} \fr{\<\phi(p)|\Ecal(\hat{n}_1)|\Ocal(p')\>\cdot\Pi_\Ocal(p')\cdot\<\Ocal(p')|\Ecal(\hat{n}_2)|\phi(p)\>}{\<\phi(p)|\phi(p)\> \; \<\Ecal(\hat{n}_1)\> \; \<\Ecal(\hat{n}_2)\>}.
\label{eq:SumStates_intro}
\ee
Here $\Pi_\Ocal(p')$ is a specific projector that we will define explicitly later. Multi-point positivity therefore provides an efficient way of accessing such constraints. 
Concretely, unitarity and positivity limit the partial wave coefficients in~\eqref{eq:PartialWave4d} to the range~\cite{Fox:1978vw,sasha_positive},\footnote{Remarkably, these bounds were derived in the early QCD literature using expressions for the energy correlators in terms of particle states, and apply beyond the case of conformal theories and the conformal block decomposition, as considered here. We thank Sasha Zhiboedov for bringing these earlier works to our attention, and emphasizing the importance of this constraint \cite{sasha_positive}.}
\be
0 \leq \Fcal_\Ecal^{(\ell)} \leq 1.
\ee
Combining these positivity constraints with the conformal block expansion~\eqref{eq:ConfBlock4d} leads to nontrivial bounds on the OPE coefficients $\lambda_{\phi T\Ocal}$, which must be satisfied in any CFT. In contrast with typical bootstrap bounds, these constraints do not directly rely on crossing symmetry and instead follow simply from energy positivity. However, it would be interesting to understand if these bounds can be directly proven from the bootstrap of local correlators. As a proof-of-principle example, we apply these ideas to the three-dimensional Ising CFT, obtaining new bounds on both parity-even and parity-odd couplings to the stress tensor.

Altogether, our results provide the first general construction of conformal blocks for energy correlators and establish a non-perturbative framework for understanding the full operator content of the energy-energy correlator. This framework offers a unified perspective that connects collider-inspired observables, conformal symmetry, and positivity, and sets the stage for future bootstrap applications and non-perturbative studies of energy flow in quantum field theories.

The outline of this paper is as follows. In Sec.~\ref{sec:Method} we review the approach to the calculation of the t-channel conformal blocks for the energy-energy correlator, and set up our notation. In Sec.~\ref{sec:Blocks} we compute the energy correlator conformal block for any traceless symmetric operator $\Ocal_{\mu_1\cdots\mu_J}$ appearing in the OPE of the source $\phi$ with the stress tensor. In Sec.~\ref{sec:ProductCFT} we perform a simple study of the convergence within the context of a tensor product CFT. In Sec.~\ref{sec:Bounds} we derive new bounds on OPE coefficients of parity-even and parity-odd OPE coefficients in the 3d Ising CFT. We conclude in Sec.~\ref{sec:Discussion} and discuss a number of future directions.


\section{Method and Notation}
\label{sec:Method}

In this section, we define all the ingredients in our calculation of energy correlator conformal blocks for general spacetime dimension $d$.\footnote{In this work we use the ``mostly minus'' metric $\eta_{\mu\nu} = \textrm{diag}(+,-,\ldots,-)$.} To start, we need to first more carefully define the energy operator $\Ecal(\hat{n})$. As discussed in~\cite{Belitsky:2013xxa}, in a CFT all energy emitted by an external source propagates to future null infinity, so the limit $r\ra\infty$ should be taken with the retarded time $t - r$ fixed. We can ensure this by rewriting the energy operator in the more covariant form\footnote{This definition of $\Ecal(n)$ is related to the light transform of $T_{\mu\nu}$ defined in~\cite{Kravchuk:2018htv} by an overall factor of 2 (see Sec.~2.3 of~\cite{Korchemsky:2021okt}),
\be
\mathbf{L}[T](\infty;n) \equiv \lim_{x\ra\infty} \int_{-\infty}^\infty d\alpha \; (-\alpha)^{-d-2} \; T\Big(x-\fr{n}{\alpha};n\Big) = \half\Ecal(n).
\ee}
\be
\Ecal(n) \equiv \fr{1}{(n\cdot\bar{n})^2} \int_{-\infty}^\infty d(x\cdot n) \lim_{x\cdot\bar{n}\ra\infty} \bigg(\fr{x\cdot\bar{n}}{n\cdot\bar{n}}\bigg)^{d-2} T(x;\bar{n}),
\label{eq:Edef}
\ee
where $n^\mu \equiv (1,\hat{n})$ is the null vector labeling the direction of energy detection and $\bar{n}^\mu$ is an arbitrary null vector satisfying $n\cdot\bar{n} \neq 0$. Note that it is common to choose $\bar{n}^\mu = (1,-\hat{n})$, such that $x\cdot\bar{n}$ is the advanced time $t + r$. However, as we will see in the following sections, all correlation functions of $\Ecal(n)$ are independent of our choice of $\bar{n}$. In writing~\eqref{eq:Edef} we have used the index-free notation
\be
\Ocal(x;n) \equiv n^{\mu_1}\cdots n^{\mu_J} \Ocal_{\mu_1\cdots\mu_J}.
\ee

In this work, we are specifically interested in the two-point function of the energy operator, as defined in~\eqref{eq:ECorrDef}, in the external state created by a scalar primary operator $\phi$ inserted with timelike momentum $p^\mu$,
\be
|\psi\> = |\phi(p)\> \equiv \int d^dx \; e^{-ip\cdot x} \phi(x)|0\>.
\ee
It will be convenient to normalize this energy correlator by the one-point function, which is fixed by conformal invariance to
\be
\<\Ecal(n)\> = \fr{\G(\fr{d-1}{2})}{2\pi^{\fr{d-1}{2}}} \; \fr{p^{d}}{(p\cdot n)^{d-1}}.
\ee
We can therefore define the dimensionless function
\be
\Fcal_\Ecal(z) \equiv \fr{\<\Ecal(n_1) \; \Ecal(n_2)\>}{\<\Ecal(n_1)\> \; \<\Ecal(n_2)\>},
\label{eq:FDef}
\ee
which depends only on the cross-ratio
\be
z \equiv \fr{p^2(n_1\cdot n_2)}{2(p\cdot n_1) (p\cdot n_2)}.
\label{eq:CrossRatio}
\ee
If we boost to the rest frame of the external source, this cross-ratio simply measures the angle $\theta$ between the two detectors,
\be
p^\mu = (E,\vec{0}) \, \ra \, z = \half\big(1-\cos\theta\big).
\ee
One useful representation of this two-point function is to decompose it in terms of $SO(d-1)$ partial waves, labeled by the integer $\ell$,
\be
\Fcal_\Ecal(z) = \sum_{\ell=0}^\infty \Fcal_\Ecal^{(\ell)} \; \Chat_\ell^{(\fr{d-3}{2})}(1-2z),
\label{eq:FPartialWave}
\ee
where $\Chat_\ell^{(\fr{d-3}{2})}(x)$ is a rescaled Gegenbauer polynomial,
\be
\Chat_\ell^{(\fr{d-3}{2})}(x) \equiv \bigg(\fr{2\ell+d-3}{d-3}\bigg) C_\ell^{(\fr{d-3}{2})}(x),
\ee
which has been normalized such that the partial wave coefficients are finite for $d\ra3$.\footnote{We can see this explicitly by writing these polynomials in the general form
\be
\Chat_\ell^{(\fr{d-3}{2})}(x) = \bigg(\fr{2\ell+d-3}{\ell+d-3}\bigg) \fr{(d-2)_\ell}{\ell!} \; {}_2F_1\Big(\!-\ell,\ell+d-3;\fr{d-2}{2};\fr{1-x}{2}\Big),
\ee
which has the limit
\be
\lim_{d\ra3} \Chat_\ell^{(\fr{d-3}{2})}(\cos\theta) = \fr{2\cos(\ell\theta)}{(1+\de_{\ell0})}.
\ee}

In a CFT, the energy two-point function can also be decomposed in terms of irreducible representations of the conformal group $SO(d,2)$, each associated with a primary operator $\Ocal$ appearing in the OPE $\phi \times T_{\mu\nu}$,
\be
\Fcal_\Ecal(z) = \sum_\Ocal \lambda^2_{\phi T\Ocal} \; \Gcal_\Ocal(z),
\label{eq:ConfBlockExp}
\ee
where $\lambda_{\phi T\Ocal}$ are the OPE coefficients and $\Gcal_\Ocal(z)$ is the contribution of a single irreducible representation i.e.~the primary operator $\Ocal$ and all its descendants $\sim \p^n\Ocal$. These functions are known as \emph{conformal blocks} or conformal partial waves.\footnote{In the context of energy correlators, the blocks appearing in the s-channel OPE are sometimes referred to as ``celestial blocks" \cite{Kologlu:2019mfz}. For the t-channel OPE considered here, which directly relates to local OPE data, we will use the standard terminology of ``conformal blocks".} The individual conformal blocks are completely fixed by conformal symmetry, depending only on the scaling dimension $\De$ and Lorentz representation of the associated primary operator $\Ocal$ together with the corresponding data of the external operators $\phi$ and $T_{\mu\nu}$.
The allowed spin quantum numbers for operators in this OPE depend on the spacetime dimension $d$. In this work, we focus specifically on the conformal blocks for traceless symmetric operators with integer spin $J$, which appear for any $d$.

The conformal block decomposition~\eqref{eq:ConfBlockExp} can be derived by inserting a complete set of states in the original energy correlator, which by the operator-state correspondence can be written as a sum over primary operators (see e.g.~\cite{Gillioz:2016jnn}),
\be
\Fcal_\Ecal(z) = \sum_\Ocal \int\fr{d^dp'}{(2\pi)^d} \fr{\<\phi(p)|\Ecal(n_1)|\Ocal(p')\>\cdot\Pi_\Ocal(p')\cdot\<\Ocal(p')|\Ecal(n_2)|\phi(p)\>}{\<\phi(p)|\phi(p)\> \; \<\Ecal(n_1)\> \; \<\Ecal(n_2)\>},
\label{eq:SumStates}
\ee
where $\Pi_\Ocal(p)$ is a projection tensor ensuring we sum over all orthonormal intermediate states built from the various polarizations of $\Ocal$.

\begin{figure}[t!]
\centering
\includegraphics[width=.8\linewidth]{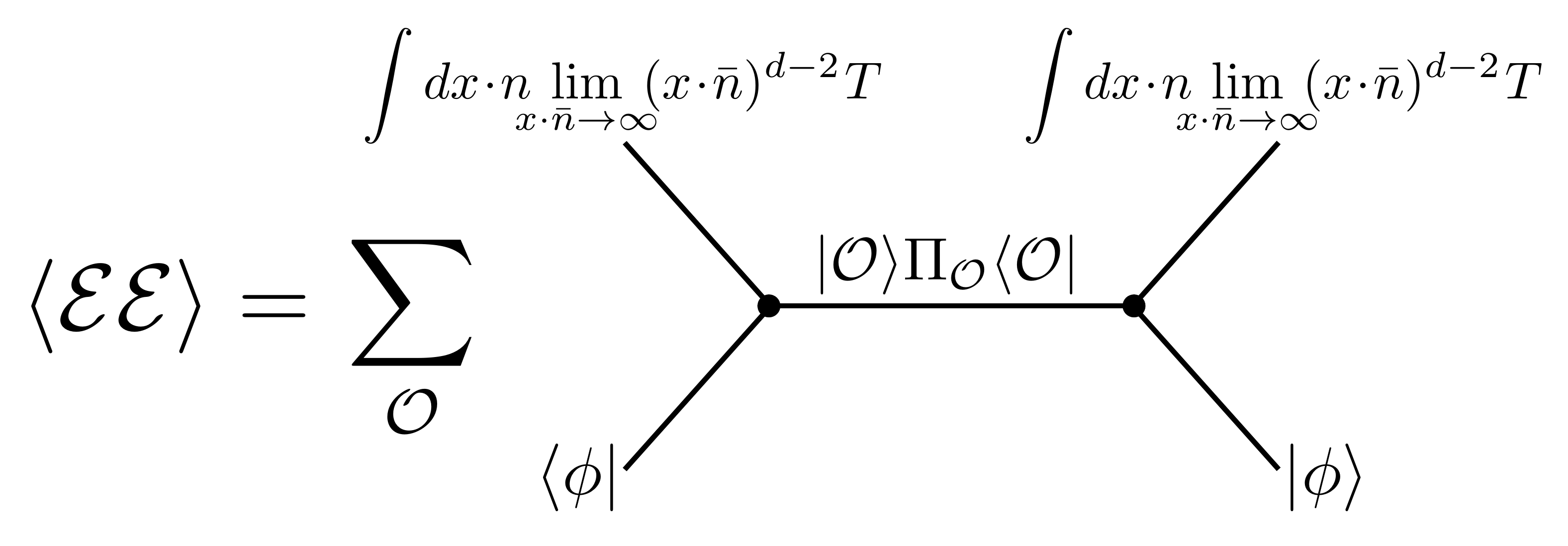}
\caption{Schematic representation of the conformal block decomposition of the two-point energy correlator~\eqref{eq:SumStates}. Each conformal block is constructed from three-point functions of the external operator $\phi$, the energy operator $\Ecal$ obtained from $T_{\mu\nu}$ via~\eqref{eq:Edef}, and the exchanged primary operator $\Ocal$.}
\label{fig:ConformalBlocks}
\end{figure}

As shown schematically in Figure~\ref{fig:ConformalBlocks}, each conformal block is therefore built from a product of three-point functions, which are glued together by the projection tensor $\Pi_\Ocal(p)$. Following the general procedure presented in~\cite{Kologlu:2019bco}, our strategy for computing these blocks for traceless symmetric operators is to decompose the projection tensor $\Pi_\Ocal$ in the basis of projectors from the spin-$J$ representation of $SO(d-1,1)$ to the spin-$\ell$ representations of $SO(d-1)$. Using index-free notation, this decomposition takes the form
\be
\ba
\Pi_\Ocal(p;n_1,n_2) &\equiv n_{1\mu_1}\cdots n_{1\mu_J} n_{2\nu_1} \cdots n_{2\nu_J} \Pi_\Ocal^{\mu_1\cdots\mu_J;\nu_1\cdots\nu_J}(p) \\
&= \fr{(p\cdot n_1)^J (p\cdot n_2)^J}{p^{2\De+2J-d}} \sum_{\ell=0}^J P_\Ocal^{(\ell)} \; \Pi_\ell^J(z),
\ea
\ee
where $z$ is the same cross-ratio from~\eqref{eq:CrossRatio}. Unsurprisingly, this basis of $SO(d-1)$ representations is constructed from the same Gegenbauer polynomials above,\footnote{Note that our definition of $\Pi_\ell^J(z)$ differs from that of~\cite{Kologlu:2019bco} by a factor of $(-1)^\ell$, in order to make the positivity of the resulting conformal blocks manifest.}
\be
\Pi_\ell^J(z) = \fr{J!\;(d-2+J+\ell)_{J-\ell}}{2^J\;(J-\ell)!\;(\fr{d-1}{2})_J} \; \Chat_\ell^{(\fr{d-3}{2})}(1-2z),
\label{eq:PiJLdef}
\ee
where $(x)_n \equiv \fr{\G(x+n)}{\G(x)}$ is the rising Pochhammer symbol.

We can determine the coefficients $P_\Ocal^{(\ell)}$ by inserting this projector in the two-point function of $\Ocal$ itself,
\be
\<\Ocal(p;n_1)|\Ocal(p';n_2)\> = \int\fr{d^dq}{(2\pi)^d} \; \<\Ocal(p;n_1)|\Ocal_{\mu_1\cdots\mu_J}(q)\>\;\Pi_\Ocal^{\mu_1\cdots\mu_J;\nu_1\cdots\nu_J}(q)\;\<\Ocal_{\nu_1\cdots\nu_J}(q)|\Ocal(p';n_2)\>.
\label{eq:TwoPtProj}
\ee
Decomposing this two-point function also in terms of $SO(d-1)$ representations,
\be
\<\Ocal(p;n_1)|\Ocal(p';n_2)\> \equiv (2\pi)^d\de^d(p-p') \; \Theta(p_0) \; \Theta(p^2) \; p^{2\tau-d} (p\cdot n_1)^J (p\cdot n_2)^J \sum_{\ell=0}^J F_\Ocal^{(\ell)} \; \Pi_\ell^J(z),
\ee
where $\tau \equiv \De-J$ is the so-called ``twist'' of $\Ocal$ and $\Theta(x)$ is the Heaviside step function, we can rewrite~\eqref{eq:TwoPtProj} as
\be
\sum_{\ell=0}^J F_\Ocal^{(\ell)} \; \Pi_\ell^J(z) = \sum_{\ell=0}^J \big(F_\Ocal^{(\ell)}\big)^2 \; P_\Ocal^{(\ell)} \; \Pi_\ell^J(z),
\ee
which fixes the projector coefficients to be the inverse of the two-point function coefficients,
\be
P_\Ocal^{(\ell)} = \fr{1}{F_\Ocal^{(\ell)}}.
\ee

If we similarly decompose the three-point functions in~\eqref{eq:SumStates} into $SO(d-1)$ representations,
\be
\ba
&\<\phi(p)|\Ecal(n_1)|\Ocal(p';n_2)\> \\
&\qquad \equiv (2\pi)^d\de^d(p-p') \; \Theta(p_0) \; \Theta(p^2) \; \lambda_{\phi T\Ocal} \; \<\Ecal(n_1)\> \; p^{\tau+\De_\phi-d} (p\cdot n_2)^J \sum_{\ell=0}^J \Ccal_{\phi\Ecal\Ocal}^{(\ell)} \; \Pi_\ell^J(z),
\ea
\ee
we can write each conformal block as a sum over partial waves,
\be
\Gcal_\Ocal(z) \equiv \sum_{\ell=0}^J \Gcal_\Ocal^{(\ell)} \; \Chat_\ell^{(\fr{d-3}{2})}(z),
\ee
with the coefficients
\be
\boxed{\Gcal_\Ocal^{(\ell)} = \fr{J!\;(d-2+J+\ell)_{J-\ell}}{2^J\;(J-\ell)!\;(\fr{d-1}{2})_J} \; \fr{|\Ccal_{\phi\Ecal\Ocal}^{(\ell)}|^2}{F_\phi \; F_\Ocal^{(\ell)}}.}
\label{eq:Block}
\ee
To obtain the conformal blocks, we therefore just need to compute the partial wave coefficients of the two-point function of $\Ocal$ and the three-point function $\<\phi|\Ecal|\Ocal\>$. The partial wave coefficients of the full energy correlator are then given by the sum
\be
\Fcal_\Ecal^{(\ell)} = \sum_\Ocal \lambda_{\phi T\Ocal}^2 \; \Gcal_\Ocal^{(\ell)}.
\ee
As we can see from~\eqref{eq:Block}, these partial wave coefficients $\Gcal_\Ocal^{(\ell)}$ are manifestly positive due to unitarity. The decomposition of $\Fcal^{(\ell)}_\Ecal$ into t-channel conformal blocks is therefore a sum of positive terms, which will be crucial in deriving bounds on OPE coefficients in Sec.~\ref{sec:Bounds}.


\section{Calculation of Conformal Blocks}
\label{sec:Blocks}

In this section, we compute the energy correlator conformal block for any traceless symmetric operator $\Ocal_{\mu_1\cdots\mu_J}$ appearing in the OPE of the source $\phi$ with the stress tensor $T_{\mu\nu}$, by following the procedure summarized in Sec.~\ref{sec:Method}.  The majority of this section applies to general CFTs in $d\geq4$, where the only allowed tensor structures are parity-even, however in the special case of $d=3$ there are additional parity-odd tensor structures, which we discuss in Sec.~\ref{sec:ParityOdd}. In order to streamline the presentation here, we have relegated various details of the calculation to App.~\ref{app:Details}.


\subsection{Two-Point Functions}
\label{sec:TwoPt}

As explained in Sec.~\ref{sec:Method}, to construct the conformal block associated with a given operator $\Ocal$ we first need to orthonormalize the intermediate states built from the possible polarizations of $\Ocal$. To do this, we need to compute the two-point function
\be
\<\Ocal(p;n_1)|\Ocal(p';n_2)\> \equiv (2\pi)^d\de^d(p-p') \; \Theta(p_0) \; \Theta(p^2) \; p^{2\tau-d} (p\cdot n_1)^J (p\cdot n_2)^J \; F_\Ocal(z),
\label{eq:TwoPtDef}
\ee
where we've used scale, translation, and rotation invariance to reduce this correlator to a function of the cross-ratio $z$ defined in~\eqref{eq:CrossRatio}. We can compute $F_\Ocal(z)$ by taking the Fourier transform of the position space two-point function, which for a general traceless symmetric operator of spin $J$ takes the form~\cite{Costa:2011mg},\footnote{The overall phase $(-1)^{-\De}$ arises from Wick rotation and ensures the resulting momentum space norm is positive.}
\be
\<\Ocal(x;n_1) \; \Ocal(0;n_2)\> = (-1)^{-\De} \fr{1}{x^{2(\De+J)}} \Big( x^2 (n_1 \cdot n_2) - 2(x \cdot n_1) (x \cdot n_2) \Big)^J.
\label{eq:TwoPtPos}
\ee
To obtain a well-defined initial and final state in~\eqref{eq:TwoPtDef}, we specifically need the Wightman two-point function, which we can obtain with the $i\epsilon$-prescription (see e.g.~\cite{Hartman:2015lfa})
\be
x^2 = (t-i\epsilon)^2 - |\vec{x}|^2.
\ee

We can evaluate the Fourier transform of this two-point function by expanding it as a sum of integrals of the general form (using the $i\epsilon$-prescription above),\footnote{See App.~B of~\cite{Firat:2023lbp} for the evaluation of integrals of this form.}
\be
\ba
&\int d^dx \; e^{i p \cdot x} \fr{(x\cdot n_1)^\alpha \; (x\cdot n_2)^\beta}{x^{2\de}} \\
&\qquad = \Theta(p_0) \; \Theta(p^2) \; \fr{(p\cdot n_1)^\alpha \; (p\cdot n_2)^\beta}{p^{2(\alpha+\beta-\de)+d}} \; \fr{(-1)^{\de-\half(\alpha+\beta)} \;\pi^{\fr{d+2}{2}} \; {}_2F_1(-\alpha,-\beta;\de-\alpha-\beta-\fr{d-2}{2};z)}{2^{2\de-\alpha-\beta-d-1}\G(\de) \; \G(\de-\alpha-\beta-\fr{d-2}{2})},
\ea
\label{eq:BBFourier}
\ee
with the resulting expression
\be
F_{\Ocal}(z) = \fr{\pi^{\fr{d+2}{2}} \; {}_2F_1(-J,\De-1;\tau-\fr{d-2}{2};z)}{2^{2\De-J-d-1} \; \G(\De+J) \; \G(\tau-\fr{d-2}{2})}.
\label{eq:TwoPtMom}
\ee
Due to the non-positive integer in the first entry of the hypergeometric function, the two-point function is therefore a degree-$J$ polynomial in $z$.

We then need to decompose this polynomial into the basis of $SO(d-1)$ representations $\Pi^J_\ell(z)$ defined in~\eqref{eq:PiJLdef},
\be
F_\Ocal(z) = \sum_{\ell=0}^J F_\Ocal^{(\ell)} \; \Pi_\ell^J(z),
\ee
which we can do by evaluating integrals of the form
\be
\int_0^1 \! \fr{dz}{\big(z(1-z)\big)^{2-\fr{d}{2}}} \; z^m \; \Chat^{(\fr{d-3}{2})}_\ell(1-2z) = \fr{(-1)^\ell \; m! \; (2\ell+d-3) \; \G(\fr{d-2}{2}) \; \G(\fr{d-2}{2}+m) \; (d-2)_{\ell-1}}{\ell!\;(m-\ell)! \; \G(\ell+d-2+m)}.
\label{eq:BBGegenbauer}
\ee
We thus obtain the general two-point function coefficients
\be
\boxed{F_\Ocal^{(\ell)} = \fr{\pi^{\fr{d+2}{2}} (\De-1)_\ell \; \G(\De-\ell-d+2)}{2^{2\De-d-1} \G(\De-\fr{d-2}{2}) \; \G(\De+J) \; \G(\tau-d+2)}.}
\label{eq:TwoPtCoeffs}
\ee

As a simple example, we can consider the two-point function of the stress-energy tensor, whose position space two-point function is typically normalized with an overall factor of $c_T$ with respect to~\eqref{eq:TwoPtDef}. From~\eqref{eq:TwoPtMom} we have the resulting momentum space correlator
\be
F_T(z) = \fr{c_T \; \pi^{\fr{d+2}{2}}}{2^{d-3}(d+1) \; \G(d+2) \; \G(\fr{d}{2})} \; \Chat_2^{(\fr{d-3}{2})}(1-2z).
\label{eq:TNorm}
\ee
In other words, the only nonzero contribution comes from the $\ell=2$ representation,
\be
F_T^{(0)} = F_T^{(1)} = 0, \quad F_T^{(2)} = \fr{c_T \; \pi^{\fr{d+2}{2}}(d-1)}{2^{d-2} \G(d+2) \; \G(\fr{d}{2})},
\ee
which simply follows from the fact that the stress tensor is conserved and therefore transforms in a short multiplet of $SO(d-1,1)$.


\subsection{Three-Point Functions}
\label{sec:ThreePt}

Next we need to compute the energy operator matrix elements between the external source $\phi$ and the exchanged operator $\Ocal$,
\be
\<\phi(p)|\Ecal(n_1)|\Ocal(p';n_2)\> \equiv (2\pi)^d\de^d(p-p') \; \Theta(p_0) \; \Theta(p^2) \; \lambda_{\phi T\Ocal} \; \<\Ecal(n_1)\> \; p^{\tau+\De_\phi-d} (p\cdot n_2)^J \; \Ccal_{\phi\Ecal\Ocal}(z).
\ee
We can obtain $\Ccal_{\phi\Ecal\Ocal}(z)$ by starting with the position space three-point function $\<\phi T \Ocal\>$, using~\eqref{eq:Edef} to obtain a position space correlator with $\Ecal$, then Fourier transforming the external states to momentum space.

In general, CFT three-point functions involving multiple operators with spin can have more than one independent tensor structure, each of which has its own theory-dependent OPE coefficient. In $d\geq4$ all possible tensor structures for $\<\phi T \Ocal\>$ can be constructed from the parity-even building blocks~\cite{Costa:2011mg}
\be
\ba
H_{ij} &\equiv x_{ij}^2 (n_i \cdot n_j) - 2 (x_{ij} \cdot n_i) (x_{ij} \cdot n_j), \\
V_{i,jk} &\equiv \fr{x_{ij}^2(x_{ik} \cdot n_i) - x_{ik}^2 (x_{ij} \cdot n_i)}{x_{jk}^2}.
\ea
\label{eq:HandV}
\ee
In $d=3$ there are additional parity-odd tensor structures, which we discuss in Sec.~\ref{sec:ParityOdd}. Using these building blocks, we can construct the general three-point function
\be
\ba
&\<\phi(x_0) \; T(x_1;n_1) \; \Ocal(x_2;n_2)\> \\
& \qquad = (-1)^{-\half(\De_\phi+\De+d)}\fr{\lambda^{(0)}_{\phi T\Ocal} V_{1,02}^2 \; V_{2,01}^J + \lambda^{(1)}_{\phi T\Ocal} V_{1,02} \; V_{2,01}^{J-1} \; H_{12} + \lambda^{(2)}_{\phi T\Ocal} V_{2,01}^{J-2} \; H_{12}^2}{x_{01}^{\De_\phi-\De-J+d+2} x_{12}^{\De+J-\De_\phi+d+2} x_{02}^{\De_\phi+\De+J-d-2}}.
\ea
\label{eq:ThreePtPos}
\ee
Just like for the two-point function, we are specifically interested in the Wightman function with the ordering shown here, which we can ensure by analytically continuing $t_i \ra t_i-i\ep_i$ with $\ep_0 > \ep_1 > \ep_2$.

Naively, we therefore have three independent OPE coefficients for each operator $\Ocal$. However, conservation of the stress tensor imposes the additional constraints~\cite{Meltzer:2018tnm}
\be
\ba
\lambda^{(1)}_{\phi T\Ocal} &= -\fr{2\big( (d-1)(\De-\De_\phi)+J\big)}{(d-2)(\tau-\De_\phi-d)} \; \lambda^{(0)}_{\phi T\Ocal}, \\
\lambda^{(2)}_{\phi T\Ocal} &= \fr{(d-1)(\De-\De_\phi)^2-(d-2)J-J^2}{(d-2)(\tau-\De_\phi-d)(\tau-\De_\phi-d+2)} \; \lambda^{(0)}_{\phi T\Ocal}.
\ea
\label{eq:WardOPE}
\ee
The entire three-point function is therefore completely fixed by conformal symmetry up to one overall coefficient $\lambda_{\phi T \Ocal} \equiv \lambda^{(0)}_{\phi T \Ocal}$. One consequence of these relations between OPE coefficients is that $\lambda_{\phi T\Ocal} = 0$ if $\Ocal$ is a scalar operator with $\De \neq \De_\phi$ or a vector operator with $\De \neq \De_\phi \pm 1$~\cite{Giombi:2011rz,Meltzer:2017rtf}.

From this position space correlator, we can obtain the resulting momentum space three-point function (see App.~\ref{app:Details} for a detailed derivation)
\be
\boxed{
\ba
\Ccal_{\phi\Ecal\Ocal}(z) &= \fr{(-1)^{\half(\De_\phi-\tau)} \; \pi^{d+\fr{3}{2}} \; \G(d+1)}{2^{\De_\phi+\De-2} \; \G(\fr{d-1}{2}) \; \G(\fr{\De_\phi-\tau+d+2}{2})} \Bigg( \fr{{}_3F_2\big(\!-\!J,d-1,\De+1;\fr{\tau-\De_\phi+d+2}{2},\fr{\tau+\De_\phi+2}{2};z\big)}{\G(\fr{\tau-\De_\phi+d+2}{2}) \; \G(\fr{\tau+\De_\phi+2}{2}) \; \G(\fr{\De_\phi+\De+J-d+2}{2})} \\
&\qquad + \; \fr{4(d-1)(\fr{\De_\phi-\De-J}{2})}{(d-2)} \fr{z \; {}_3F_2\big(1-J,d,\De+1;\fr{\tau-\De_\phi+d+4}{2},\fr{\tau+\De_\phi+4}{2};z\big)}{\G(\fr{\tau-\De_\phi+d+4}{2}) \; \G(\fr{\tau+\De_\phi+4}{2}) \; \G(\fr{\De_\phi+\De+J-d}{2})} \\
&\qquad + \; \fr{4(d-1)(\fr{\De_\phi-\De-J}{2})_2}{(d-2)} \fr{z^2 \; {}_3F_2\big(2-J,d+1,\De+1;\fr{\tau-\De_\phi+d+6}{2},\fr{\tau+\De_\phi+6}{2};z\big)}{\G(\fr{\tau-\De_\phi+d+6}{2}) \; \G(\fr{\tau+\De_\phi+6}{2}) \; \G(\fr{\De_\phi+\De+J-d-2}{2})} \Bigg).
\ea
}
\label{eq:ThreePtMom}
\ee
While this expression may seem rather complicated, it's simply a degree-$J$ polynomial in $z$ (albeit with involved coefficients).


\subsection{Partial Wave Coefficients and Ward Identities}
\label{sec:PartialWave}

The final ingredient needed to obtain the conformal blocks is the decomposition of the three-point function~\eqref{eq:ThreePtMom} in terms of $SO(d-1)$ representations,
\be
\Ccal_{\phi\Ecal\Ocal}(z) = \sum_{\ell=0}^J \Ccal_{\phi\Ecal\Ocal}^{(\ell)} \; \Pi_\ell^J(z),
\ee
which we can again obtain from integrals of the form~\eqref{eq:BBGegenbauer} (see App.~\ref{app:Details} for details).

These coefficients have unique properties for $\ell=0$ and $\ell=1$, so we'll consider those cases individually before proceeding to the more general case $\ell \geq 2$. Starting with $\ell=0$, we find\footnote{As mentioned in the previous section, Ward identities restrict the allowed scaling dimensions of $\Ocal$ for $J=0,1$, but here we have kept $\De$ general.}
\be
\Ccal_{\phi\Ecal\Ocal}^{(0)} = \begin{cases} \ba &\fr{(-1)^{\half(\De_\phi-\De)} \; 2^{2-\De_\phi-\De} \; \pi^{d+\fr{3}{2}} \; \G(d+1)}{\G(\fr{d-1}{2}) \; \G(\fr{\De_\phi-\De+d+2}{2}) \; \G(\fr{\De_\phi+\De-d+2}{2}) \; \G(\fr{\De-\De_\phi+d+2}{2}) \; \G(\fr{\De_\phi+\De+2}{2})} & J=0, \\ &\fr{(-1)^{\half(\De_\phi-\tau)} \pi^{d+\fr{3}{2}} \big(\De_\phi \; d(\De_\phi-d)-(\De+1)(d(\De-d-1)+4)\big) \; \G(d+1)}{2^{\De_\phi+\De}(d-2) \; \G(\fr{d-1}{2}) \; \G(\fr{\De_\phi-\De+d+3}{2}) \; \G(\fr{\De_\phi+\De-d+3}{2}) \; \G(\fr{\De-\De_\phi+d+3}{2}) \; \G(\fr{\De_\phi+\De+3}{2})} & J=1, \\ &0 & J \geq 2. \ea \end{cases}
\label{eq:ThreePtL0}
\ee
From this we see that the only conformal blocks which can contribute to $\Fcal_\Ecal^{(0)}$ are those for scalars and vectors. This restriction is important, as the $\ell=0$ coefficient simply measures the total energy of the external state and is therefore fixed to be~\cite{Basham:1978bw}
\be
\Fcal_\Ecal^{(0)} = 1.
\label{eq:WardL0}
\ee

One operator which we know must appear in the $\phi \times T_{\mu\nu}$ OPE is $\phi$ itself. Using~\eqref{eq:Block} we can obtain the associated conformal block
\be
\Gcal_\phi = \Gcal_\phi^{(0)} = \fr{|\Ccal_{\phi\Ecal\phi}^{(0)}|^2}{(F_\phi)^2} = \fr{(d-1)^2\pi^d}{\De_\phi^2 \; \G^2(\fr{d+2}{2})}.
\ee
The associated OPE coefficient is fixed by Ward identities to~\cite{Osborn:1993cr}
\be
\lambda_{\phi T\phi} = \fr{\De_\phi \; \G(\fr{d+2}{2})}{(d-1)\pi^{\fr{d}{2}}},
\ee
such that we obtain the universal contribution
\be
\lambda_{\phi T\phi}^2 \; \Gcal_\phi^{(0)} = 1.
\ee
In other words, the $\phi$ conformal block automatically \emph{saturates} the energy conservation requirement~\eqref{eq:WardL0}. Because all other potential conformal block contributions are necessarily positive, this means that \emph{no other operators} can contribute to $\Fcal_\Ecal^{(0)}$. We therefore rediscover the well-known result that $\lambda_{\phi T\Ocal}=0$ for all scalar operators $\Ocal_{J=0} \neq \phi$, as well as the additional constraint:
\benn
\lambda_{\phi T\Ocal} = 0 \textrm{ for all vector operators } \Ocal_{J=1}.
\eenn
As mentioned in the previous section, position space Ward identities already constrain these OPE coefficients to be zero unless $\De = \De_\phi \pm 1$. It was also argued in~\cite{Bzowski:2013sza} that this OPE coefficient must vanish when $\Ocal$ is a conserved current. This new constraint generalizes those results to all vector operators, regardless of their scaling dimension.\footnote{There are examples of free theories with multiple fields where one can construct additional conserved spin-two operators that have nonzero OPE coefficients between scalars and spin-one operators~\cite{Giombi:2011rz}. The constraint we discuss here is specifically for the actual stress-energy tensor, which measures the total energy.}

Next let's consider $\ell=1$, with the associated three-point function coefficients
\be
\Ccal_{\phi\Ecal\Ocal}^{(1)} = \begin{cases} \ba &\fr{(-1)^{\half(\De_\phi-\tau)} \; 2^{1-\De_\phi-\De} \; \pi^{d+\fr{3}{2}} \; (d-1) \big(\De_\phi(d-\De_\phi)+\De^2-1\big) \; \G(d+1)}{(d-2) \; \G(\fr{d-1}{2}) \; \G(\fr{\De_\phi-\De+d+3}{2}) \; \G(\fr{\De_\phi+\De-d+3}{2}) \; \G(\fr{\De-\De_\phi+d+3}{2}) \; \G(\fr{\De_\phi+\De+3}{2})} & J=1, \\ &0 & J \geq 2. \ea \end{cases}
\label{eq:ThreePtL1}
\ee
Similarly to the $\ell=0$ case, we find that only vector operators can possibly contribute to $\Fcal_\Ecal^{(1)}$. However, we have just proven that all OPE coefficients for vector operators must be zero. We therefore rediscover the momentum conservation constraint
\be
\Fcal_\Ecal^{(1)} = 0,
\label{eq:WardL1}
\ee
which holds in all CFTs (but not more general gapped QFTs)~\cite{Kologlu:2019mfz,Korchemsky:2019nzm}.

Finally, we can turn to the remaining three-point function coefficients with $\ell \geq 2$, which can be written in the general form
\be
\boxed{
\ba
&\Ccal_{\phi\Ecal\Ocal}^{(\ell\geq2)} = \fr{(-1)^{\half(\De_\phi-\tau)} \; 2^{5-\De_\phi-\De-J-d-2\ell} \; \pi^{d+2} \; \G(d+1) \; \G(d-2+J+\ell) \; (d-1)_\ell}{\G(\fr{\De_\phi-\tau+d+2}{2}) \; \G(\fr{\tau-\De_\phi+d+2}{2}+\ell) \; \G(\fr{\tau+\De_\phi+2}{2}+\ell) \; \G(\fr{d-1}{2}) \; \G(\fr{d-1}{2}+\ell) \; (\fr{d-2}{2})_J} \\
&\quad \times \Bigg( \fr{(\De+1)_\ell}{\G(\fr{\De_\phi+\De+J-d+2}{2}) \; \G(\fr{d-2}{2})} {}_4 F_3\bigg(\genfrac..{0pt}{}{\ell-J,d-1+\ell,\De+1+\ell,\fr{d-2}{2}+\ell}{\fr{\tau-\De_\phi+d+2+2\ell}{2},\fr{\tau+\De_\phi+2+2\ell}{2},d-2+2\ell}\;;1\bigg) \\
&\quad - \; \fr{2\ell(\fr{\De_\phi-\De-J}{2}) \; (\De+1)_{\ell-1}}{J \; \G(\fr{\De_\phi+\De+J-d}{2}) \; \G(\fr{d}{2})} \; {}_5 F_4\bigg(\genfrac..{0pt}{}{\ell-J,d-1+\ell,\De+\ell,\fr{d-2}{2}+\ell,\ell+1}{\fr{\tau-\De_\phi+d+2+2\ell}{2},\fr{\tau+\De_\phi+2+2\ell}{2},d-2+2\ell,\ell}\;;1\bigg) \\
&\quad+ \; \fr{\ell(\ell-1)(\fr{\De_\phi-\De-J}{2})_2 \; (\De+1)_{\ell-2}}{J(J-1) \; \G(\fr{\De_\phi+\De+J-d-2}{2}) \; \G(\fr{d+2}{2})} \; {}_5 F_4\bigg(\genfrac..{0pt}{}{\ell-J,d-1+\ell,\De-1+\ell,\fr{d-2}{2}+\ell,\ell+1}{\fr{\tau-\De_\phi+d+2+2\ell}{2},\fr{\tau+\De_\phi+2+2\ell}{2},d-2+2\ell,\ell-1}\;;1\bigg) \! \Bigg).
\ea
}
\label{eq:ThreePtL}
\ee
Because these coefficients are nonzero only for $\ell \leq J$, each hypergeometric function in this expression is a sum of a finite number of terms, which means the numerical evaluation of these coefficients can be done relatively quickly in practice.

As a concrete example, let's use these coefficients to compute the conformal block associated with exchange of the stress tensor. The only nonzero coefficient is $\ell=2$, so we can use~\eqref{eq:Block} to obtain
\be
\ba
\Gcal_T(z) &= \Gcal_T^{(2)} \; \Chat_2^{(\fr{d-3}{2})}(1-2z) = \fr{2}{d^2-1} \; \fr{|\Ccal_{\phi\Ecal T}^{(2)}|^2}{F_\phi \; F_T^{(2)}} \; \Chat_2^{(\fr{d-3}{2})}(1-2z) \\
&= \fr{d^2(d-1)^2 \; \pi^d \; \G^2(\fr{d-2}{2}) \; \G(\fr{d}{2}) \; \G(d+2) \; \G(\De_\phi) \; \G(\De_\phi-\fr{d-2}{2})}{32 c_T \; \G^4(\fr{\De_\phi}{2}+2) \; \G^2(\fr{\De_\phi+d}{2}) \; \G^2(d-\fr{\De_\phi}{2})} \big( (d-1)(1-2z)^2 - 1 \big).
\ea
\ee


\subsection{Parity-Odd Correlators}
\label{sec:ParityOdd}

So far our discussion has been restricted to correlation functions which are even under the parity transformation $\vec{x} \ra -\vec{x}$. While these are the only allowed correlators involving a scalar source in $d\geq4$, for $d=3$ there are additional three-point function tensor structures which are odd under parity~\cite{Giombi:2011rz,Costa:2011mg}. These new tensor structures can be constructed by including the parity-odd building block
\be
S_{ij,k} \equiv \fr{x_{ij}}{x_{ik} \; x_{jk}} \; \vep_{\mu\nu\rho} \Big( n_i^\mu \; n_j^\nu \big( x_{ik}^\rho \; x_{jk}^2 - x_{jk}^\rho \; x_{ik}^2 \big) + 2 x_{ik}^\mu \; x_{jk}^\nu \big( n_i^\rho(x_{jk}\cdot n_j) - n_j^\rho(x_{ik}\cdot n_i) \big) \Big),
\ee
in addition to the usual parity-even building blocks $H_{ij}$ and $V_{i,jk}$.

Using this new building block, we can construct two possible parity-odd tensor structures for the three-point function $\<\phi T \Ocal\>$ in $d=3$,
\be
\ba
\<\phi(x_0) \; T(x_1;n_1) \; \Ocal(x_2;n_2)\>^{(-)} = (-1)^{-\half(\De_\phi+\De)+1} \fr{\lambda^{(0^-)}_{\phi T\Ocal} V_{1,02} \; V_{2,01}^{J-1} \; S_{12,0} + \lambda^{(1^-)}_{\phi T\Ocal} V_{2,01}^{J-2} \; H_{12} \; S_{12,0}}{x_{01}^{\De_\phi-\De-J+5} x_{12}^{\De+J-\De_\phi+5} x_{02}^{\De_\phi+\De+J-5}}.
\ea
\label{eq:ThreePtPosOdd}
\ee
As in the parity-even case, conservation of the stress tensor constrains these OPE coefficients, imposing the relation
\be
\lambda_{\phi T\Ocal}^{(1^-)} = - \bigg(\fr{\De-\De_\phi}{\tau-\De_\phi-2}\bigg) \lambda_{\phi T\Ocal}^{(0^-)}.
\label{eq:WardOPEOdd}
\ee
There is therefore only one independent parity-odd OPE coefficient $\lambda_{\phi T\Ocal}^{(-)} \equiv \lambda_{\phi T\Ocal}^{(0^-)}$.

In a CFT with parity symmetry, each operator can be assigned a definite charge $\pm 1$ under parity. A given three-point function $\<\phi T \Ocal\>$ will therefore either have only even or odd tensor structures, with a single associated OPE coefficient, depending on the parity charges of $\phi$ and $\Ocal$. In a CFT which violates parity, each three-point function will generically be a linear combination of odd and even tensor structures, with two independent OPE coefficients.

The parity-odd energy operator matrix element constructed from~\eqref{eq:ThreePtPosOdd} can be written in the general form
\be
\ba
&\<\phi(p) | \Ecal(n_1) | \Ocal(p';n_2)\>^{(-)} \\
&\qquad = (2\pi)^d\de^d(p-p') \; \Theta(p_0) \; \Theta(p^2) \; \lambda_{\phi T\Ocal}^{(-)} \; \<\Ecal(n_1)\> \; \fr{p^{\tau+\De_\phi-d+1} \vep_{\mu\nu\rho} \; p^\mu n_1^\nu n_2^\rho}{(p\cdot n_1)(p\cdot n_2)^{1-J}} \; \Ccal^{(-)}_{\phi\Ecal\Ocal}(z).
\ea
\ee
Following the same procedure as in the parity-even case (see App.~\ref{app:Details} for more details), we can compute the resulting expression
\be
\boxed{
\ba
\Ccal^{(-)}_{\phi\Ecal\Ocal}(z) &= \fr{(-1)^{\half(\De_\phi-\tau)} \; 3\pi^{\fr{9}{2}}}{2^{\De_\phi+\De-3} \; \G(\fr{\De_\phi-\tau+4}{2})} \Bigg( \fr{{}_3F_2\big(1-J,3,\De+1;\fr{\tau-\De_\phi+6}{2},\fr{\tau+\De_\phi+3}{2};z\big)}{\G(\fr{\tau-\De_\phi+6}{2}) \; \G(\fr{\tau+\De_\phi+3}{2}) \; \G(\fr{\De_\phi+\De+J-2}{2})} \\
&\qquad + \; \fr{2\big(\fr{\De_\phi-\De-J+1}{2}\big)\; z \; {}_3F_2\big(2-J,4,\De+1;\fr{\tau-\De_\phi+8}{2},\fr{\tau+\De_\phi+5}{2};z\big)}{\G(\fr{\tau-\De_\phi+8}{2}) \; \G(\fr{\tau+\De_\phi+5}{2}) \; \G(\fr{\De_\phi+\De+J-4}{2})} \Bigg).
\ea
}
\label{eq:ThreePtMomOdd}
\ee

We then need to decompose this three-point function in terms of parity-odd $SO(2)$ representations,
\be
\Ccal^{(-)}_{\phi\Ecal\Ocal}(z) = \sum_{\ell=1}^J \Ccal_{\phi\Ecal\Ocal}^{(\ell^-)} \; \Pi_{\ell^-}^J(z),
\ee
which consist of the polynomials
\be
\Pi^J_{\ell^-}(z) = \fr{(2J)! \; \ell}{2^{J-1}(J-\ell)! \; (J+\ell)!} \; {}_2F_1(1-\ell,\ell+1;\tfr{3}{2};z).
\ee
Just like before, only vector operators contribute to the $\ell=1$ coefficient,
\be
\Ccal_{\phi\Ecal\Ocal}^{(1^-)} = \begin{cases} \ba &\fr{(-1)^{\half(\De_\phi-\tau)} \; 3 \pi^{\fr{9}{2}}}{2^{\De_\phi+\De-3} \G(\fr{\De_\phi-\De+5}{2}) \; \G(\fr{\De-\De_\phi+5}{2}) \; \G(\fr{\De_\phi+\De-1}{2}) \; \G(\fr{\De_\phi+\De+2}{2})} & J=1, \\ &0 & J \geq 2. \ea \end{cases}
\label{eq:ThreePtL1Odd}
\ee
The momentum conservation constraint~\eqref{eq:WardL1} therefore places the same restriction on parity-odd OPE coefficients as it does for parity-even ones:
\benn
\lambda^{(-)}_{\phi T\Ocal} = 0 \textrm{ for all vector operators } \Ocal_{J=1}.
\eenn
For the remaining parity-odd coefficients with $\ell\geq2$, we find
\be
\boxed{
\ba
&\Ccal^{(\ell^-\geq2)}_{\phi\Ecal\Ocal} = \fr{(-1)^{\half(\De_\phi-\tau)} \; 3\pi^5 \; (J+\ell)! \; \ell(\ell+1)}{2^{\De_\phi+\De+J+2\ell-3} \; J \; \G(J+\half) \; \G(\fr{\De_\phi-\tau+4}{2}) \; \G(\fr{\tau-\De_\phi+4+2\ell}{2}) \; \G(\fr{\tau+\De_\phi+1+2\ell}{2})} \\
&\quad \times \Bigg( \fr{(\De+1)_{\ell-1}}{\G(\fr{\De_\phi+\De+J-2}{2})} {}_4F_3\bigg(\genfrac..{0pt}{}{\ell-J,2+\ell,\De+\ell,\ell+\half}{\fr{\tau-\De_\phi+4+2\ell}{2},\fr{\tau+\De_\phi+1+2\ell}{2},2\ell+1}\;;1\bigg) \\
&\qquad - \; \fr{2(\ell-1) \; (\fr{\De_\phi-\De-J+1}{2}) \; (\De+1)_{\ell-2}}{3(J-1) \; \G(\fr{\De_\phi+\De+J-4}{2})} \; {}_5F_4\bigg(\genfrac..{0pt}{}{\ell-J,2+\ell,\De-1+\ell,\ell+\half,\ell}{\fr{\tau-\De_\phi+4+2\ell}{2},\fr{\tau+\De_\phi+1+2\ell}{2},2\ell+1,\ell-1}\;;1\bigg) \Bigg).
\ea
}
\label{eq:ThreePtLOdd}
\ee
As the only representations which appear in the $\phi \times T_{\mu\nu}$ OPE in $d=3$ are spin-$J$ traceless symmetric operators, we now have all building blocks necessary to construct general 3d energy correlator conformal blocks.


\section{Example: Tensor Product of CFTs}
\label{sec:ProductCFT}

As a simple but instructive example, we now consider the case of a tensor product theory $\CFT_1 \otimes \CFT_2$ built from two decoupled copies of the same underlying CFT. This setup provides a useful laboratory to study the convergence of the conformal block expansion, as mixed energy correlators of the form $\<\Ecal_1 \; \Ecal_2\>$ can be computed exactly.


\subsection{Setup and OPE Coefficients}
\label{sec:Setup}

For any CFT, we can construct a new theory by combining two decoupled copies of the CFT Hilbert space,
\be
\Hcal = \Hcal_1 \otimes \Hcal_2,
\ee
with operators $\Ocal_1$ and $\Ocal_2$ which act on the individual copies. The total stress-energy tensor for this tensor product CFT corresponds to the sum of operators
\be
T^{\mu\nu} \equiv T_1^{\mu\nu} + T_2^{\mu\nu}.
\ee
Similarly, the total energy operator is a sum of individual energy operators from the two CFTs,
\be
\Ecal(n) \equiv \Ecal_1(n) + \Ecal_2(n).
\ee
We are specifically interested in correlators of these energy operators in the background of the tensor product operator $\phi_1\phi_2$, with scaling dimension $2\De_\phi$, built from two copies of the same scalar primary $\phi$ in the underlying CFT. Concretely, we will consider the mixed correlator
\be
\Fcal_{\Ecal_1\Ecal_2}(z) \equiv \fr{\<\phi_1\phi_2(p)|\Ecal_1(n_1) \; \Ecal_2(n_2)|\phi_1\phi_2(p)\>}{\<\phi_1\phi_2(p)|\phi_1\phi_2(p)\> \; \<\Ecal_1(n_1)\> \; \<\Ecal_2(n_2)\>}.
\label{eq:E1E2Def}
\ee

Because the two CFTs are decoupled, this correlation function can be computed exactly from a product of three-point functions,
\be
\ba
\Fcal_{\Ecal_1\Ecal_2}(z) &= \int d^dx \; e^{ip\cdot x} \; \fr{\<\phi_1(x) \; \Ecal_1(n_1) \; \phi_1(0)\> \; \<\phi_2(x) \; \Ecal_2(n_2) \; \phi_2(0)\>}{\<\phi_1\phi_2(p)|\phi_1\phi_2(p)\> \; \<\Ecal_1(n_1)\> \; \<\Ecal_2(n_2)\>} \\
&= \fr{2\De_\phi\;\;\G(2\De_\phi+1)\G(2\De_\phi-\fr{d-2}{2})}{\G(2\De_\phi-d+2)\;\G(2\De_\phi+\fr{d+2}{2})} \; {}_2F_1\big(d-1,d-1;2\De_\phi+\tfr{d+2}{2};z\big).
\ea
\label{eq:ProductCFTExactZ}
\ee
Following~\eqref{eq:FPartialWave}, we can decompose this correlator in terms of $SO(d-1)$ partial waves, with the resulting coefficients
\be
\ba
\Fcal_{\Ecal_1\Ecal_2}^{(\ell)} &= \fr{\ell! \; 2^{2d-6} \; \G^2(\fr{d-1}{2})}{\pi (d-3+2\ell) \; \G(d-3+\ell)} \int_0^1 \! \fr{dz}{\big(z(1-z)\big)^{2-\fr{d}{2}}} \; \Chat^{(\fr{d-3}{2})}_\ell(1-2z) \; \Fcal_{\Ecal_1\Ecal_2}(z) \\
&= \fr{(-1)^\ell \; \De_\phi\;\G(2\De_\phi-\fr{d-2}{2}) \; \G^2(d-1+\ell) \; \G(\fr{d-2}{2}+\ell)}{(2\De_\phi+1) \; \G(2\De_\phi-\fr{d-2}{2}+\ell) \; \G(\fr{d}{2}) \; \G(d-1) \; \G(d-2+2\ell)} \\
& \qquad \times {}_3F_2\big(\ell-1,\ell-1,\tfr{d-2}{2}+\ell;2\De_\phi-\tfr{d-2}{2}+\ell,d-2+2\ell;1\big).
\ea
\label{eq:ProductCFTExactL}
\ee

\begin{figure}[t!]
\centering
\includegraphics[width=.95\linewidth]{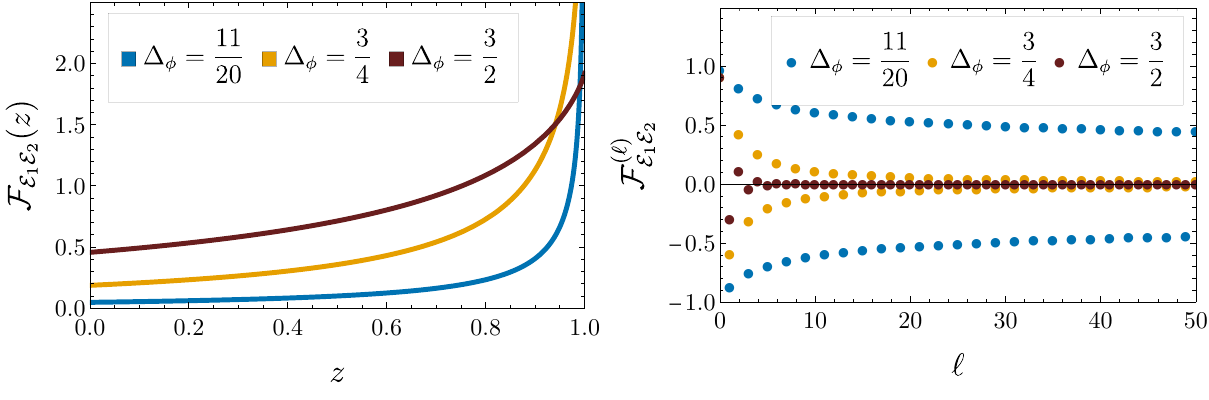}
\caption{Exact results for the mixed energy correlator $\<\Ecal_1\;\Ecal_2\>$ in a $d=3$ tensor product CFT for different values of the source scaling dimension $\De_\phi$, as a function of $z = \sin^2\fr{\theta}{2}$ (\emph{left}) and decomposed into partial waves $\Chat^{(\half)}_\ell(\cos\theta) \sim \cos\ell\theta$ (\emph{right}).}
\label{fig:ProductCFTExact}
\end{figure}

Figure~\ref{fig:ProductCFTExact} shows some examples of this mixed energy correlator, both the full function of $z$ (left) and the partial wave coefficients as a function of $\ell$ (right), for various values of $\De_\phi$ in $d=3$. As $\De_\phi$ approaches the free theory value $\fr{d-2}{2}=\half$, this correlator approaches a delta function at $\theta=\pi$, corresponding to back-to-back production of two free particles, and the associated partial wave coefficients approach $(-1)^\ell$. In the opposite limit $\De_\phi \ra \infty$, the energy correlator flattens to a homogeneous distribution~\cite{Chicherin:2023gxt,Firat:2023lbp,Cuomo:2025pjp}, with the partial wave coefficients approaching a Kronecker delta $\de_{\ell0}$.

We now want to compare this exact expression to its decomposition in terms of conformal blocks. The only operators in the $\phi_1\phi_2 \times T_i$ OPE which can contribute to this correlator are the ``double-twist'' operators of the form
\be
[\phi_1\phi_2]_{n,J} \sim \phi_1 \p^{2n} \p_{\mu_1} \cdots \p_{\mu_J} \phi_2,
\ee
with scaling dimensions $\De_{n,J} = 2\De_\phi+2n+2J$. We can therefore rewrite the mixed energy correlator as\footnote{Note that the conformal blocks in~\eqref{eq:ProductCFTExpansion} are multiplied by a factor of $4$ relative to the general blocks computed in Sec.~\ref{sec:Blocks}, due to the fact that we have chosen in~\eqref{eq:E1E2Def} to divide the mixed energy correlator by $\<\Ecal_1(n_1)\> \<\Ecal_2(n_2)\> = \fr{1}{4} \<\Ecal(n_1)\> \<\Ecal(n_2)\>$.}
\be
\Fcal_{\Ecal_1\Ecal_2}(z) = \sum_{n,J} (-1)^J \lambda_{n,J}^2 \; \Gcal_{n,J}(z),
\label{eq:ProductCFTExpansion}
\ee
where we've defined
\be
\lambda_{n,J} \equiv \lambda_{(\phi_1\phi_2) T_1 [\phi_1\phi_2]_{n,J}},
\ee
and the factor of $(-1)^J$ comes from the relation
\be
\lambda_{(\phi_1\phi_2) T_2 [\phi_1\phi_2]_{n,J}} = (-1)^J \; \lambda_{(\phi_1\phi_2) T_1 [\phi_1\phi_2]_{n,J}}.
\ee
All we need are the OPE coefficients $\lambda_{n,J}$, which can be computed via the method of ``conglomeration'' developed in~\cite{Fitzpatrick:2011dm}. In App.~\ref{app:Conglomeration}, we use this method to obtain the resulting OPE coefficients
\be
\boxed{\lambda_{n,J}^2 = \fr{\De_\phi^2 \; 2^J  \; (\fr{d-2}{2})_J^2 \; \G^2(\fr{d+2}{2})}{\pi^d \; J! \; (d-1)^2 \; (2\De_\phi+J-1)_J} \; \de_{n0},}
\label{eq:ProductCFTOPE}
\ee
such that the only conformal blocks which contribute to this correlator are those for double-twist operators with $n=0$.

The partial wave coefficients for the mixed energy correlator can therefore be written as the sum
\be
\Fcal_{\Ecal_1\Ecal_2}^{(\ell)} = \sum_{J\geq\ell} (-1)^J \lambda_J^2 \; \Gcal_J^{(\ell)},
\ee
where we've now suppressed the $n$ label on the OPE coefficients and conformal blocks, as all terms have $n=0$.


\subsection{Convergence of Partial Wave Coefficients}
\label{sec:PowerSpectrum}

Now that we have the scaling dimensions and OPE coefficients of all operators which contribute to the source-detector OPE, we can directly compare the conformal block expansion to the exact expressions for the partial wave coefficients in~\eqref{eq:ProductCFTExactL}. First we have the $\ell=0$ coefficient which, as we found in Sec.~\ref{sec:PartialWave}, only receives contributions from the spin-$0$ and spin-$1$ double-twist operators,
\be
\Fcal_{\Ecal_1\Ecal_2}^{(0)} = \lambda_0^2 \; \Gcal_0^{(0)} - \lambda_1^2 \; \Gcal_1^{(0)} = 1 - \fr{2\De_\phi-d+2}{(2\De_\phi+1)(4\De_\phi-d+2)} = \fr{2\De_\phi(4\De_\phi-d+3)}{(2\De_\phi+1)(4\De_\phi-d+2)}.
\ee
Similarly, the $\ell=1$ coefficient only receives a contribution from the $J=1$ operator,\footnote{To the attentive reader, this tensor product CFT example might naively appear to violate our earlier observation in Sec.~\ref{sec:PartialWave} that all stress tensor OPE coefficients between scalar and vector operators, and therefore the energy correlator $\ell=1$ coefficient, must be zero. This claim only holds for the \emph{total} stress-energy tensor $T \equiv T_1 + T_2$, whose associated OPE coefficients are
\be
\lambda_{(\phi_1\phi_2) T [\phi_1\phi_2]_{0,J}} = \lambda_{(\phi_1\phi_2) T_1 [\phi_1\phi_2]_{0,J}} + \lambda_{(\phi_1\phi_2) T_2 [\phi_1\phi_2]_{0,J}} = \big(1 + (-1)^J \big) \lambda_J,
\ee
which does indeed vanish for $J=1$.}
\be
\Fcal_{\Ecal_1\Ecal_2}^{(1)} = -\lambda_1^2 \; \Gcal_1^{(1)} = -\fr{2\De_\phi (d-1)}{(2\De_\phi+1)(4\De_\phi-d+2)}.
\ee

The remaining coefficients with $\ell \geq 2$ receive contributions from the infinite tower of double-twist operators with $J \geq \ell$. To study the convergence of this infinite sum we can look at the asymptotic behavior the OPE coefficients and conformal blocks as $J\ra\infty$. Starting with the OPE coefficients, we can use Stirling's approximation for gamma functions to obtain the leading behavior of~\eqref{eq:ProductCFTOPE} at large $J$,
\be
\lambda^2_J \underset{J\ra\infty}{\approx} \fr{\De_\phi^2 \; d^2(d-2)^2}{2^{2\De_\phi+2} \; \pi^{d-\half} \; (d-1)^2} \; \fr{J^{d-\fr{7}{2}}}{2^J}.
\ee
For the conformal blocks, we can use the large argument behavior of generalized hypergeometric functions~\cite{Fields:1965} to obtain the $J \ra \infty$ behavior at fixed $\ell$,
\be
\Gcal_J^{(\ell)} \underset{J\ra\infty}{\approx} \fr{2^{2\De_\phi-d+4} \; \pi^d \; (d-1)^2 \; \G^2(d-1+\ell) \; \G(2\De_\phi) \; \G(2\De_\phi-\fr{d-2}{2})}{\G^2(\ell-1) \; (d-2)^2 \; \G^2(\fr{d}{2}+1) \; \G(\fr{d-1}{2}) \; \G(2\De_\phi-d+2)} \; \fr{2^J}{J^{2\De_\phi+2d-\half}}.
\ee
Combining these together, we find the asymptotic behavior of the conformal block contributions
\be
\lambda^2_J \; \Gcal_J^{(\ell)} \underset{J\ra\infty}{\approx} \fr{2\De_\phi \; \G^2(d-1+\ell) \; \G(2\De_\phi+1) \; \G(2\De_\phi-\fr{d-2}{2})}{\G^2(\ell-1) \; \G(d-1) \; \G(\fr{d}{2}) \; \G(2\De_\phi-d+2)} \; \fr{1}{J^{2\De_\phi+d+3}}.
\ee
We therefore expect the sum over conformal blocks to \emph{converge polynomially} in $J$ for any value of $\De_\phi$ or $\ell$.

\begin{figure}[t!]
\centering
\includegraphics[width=.95\linewidth]{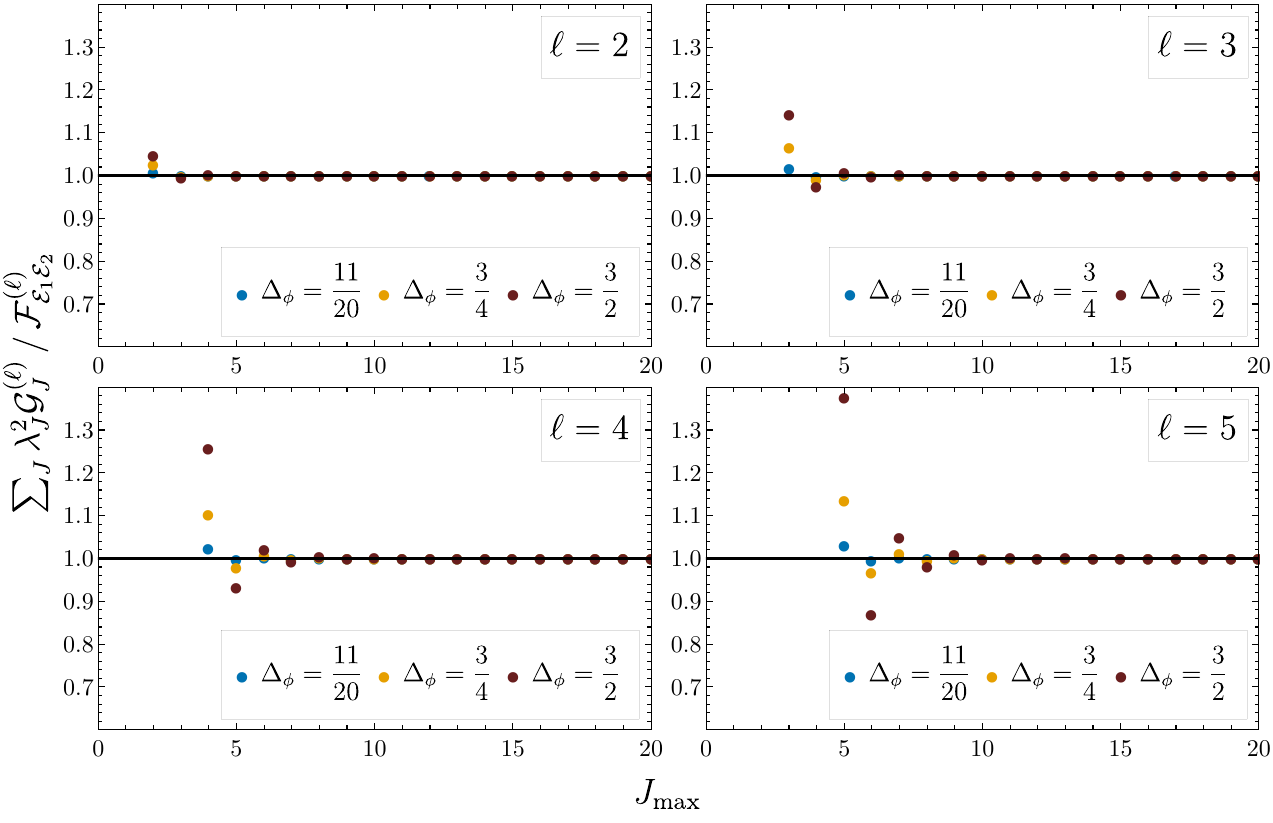}
\caption{Cumulative sum of conformal block partial wave coefficients for double-twist operators $[\phi_1\phi_2]_{0,J}$ with $J\leq J_{\max}$ as a function of $J_{\max}$ in a $d=3$ tensor product CFT, normalized by the exact energy correlator partial wave coefficient from~\eqref{eq:ProductCFTExactL}, for different values of $\ell$ and the source scaling dimension $\De_\phi$.}
\label{fig:ProductCFTConvergenceL}
\end{figure}

To test this expectation, Figure~\ref{fig:ProductCFTConvergenceL} shows the cumulative sum of double-twist conformal blocks up to a maximum spin $J_{\max}$, normalized by the exact partial wave coefficients~\eqref{eq:ProductCFTExactL}, for different values of $\De_\phi$ and $\ell$ in $d=3$. In all cases, the sum over conformal blocks converges quite rapidly, such that only the first few terms are required to obtain roughly percent-level accuracy.


\subsection{Convergence of Angular Dependence}
\label{sec:AngularDependence}

While the conformal block expansion for individual partial wave coefficients appears to converge quite rapidly, we'd also like to determine the full dependence of energy correlators on the angular parameter $z$. However, this convergence is sensitive to any singular behavior in the collinear ($z \ra 0$) and back-to-back ($z \ra 1$) limits.

For example, if we look at the exact correlator~\eqref{eq:ProductCFTExactZ} for this tensor product CFT, we see that it has a power law singularity $(1-z)^{2\De_\phi-\fr{3}{2}d+3}$ for $\De_\phi < \fr{3}{4}(d-2)$. As a result, for small $\De_\phi$ we expect this sum over conformal blocks for this correlator to only converge as a distribution. In other words, the naive sum over conformal blocks for any value of $z$ is highly oscillatory and does not converge, but the integral of this sum against a suitable test function $f(z)$ does converge.

\begin{figure}[t!]
\centering
\includegraphics[width=.95\linewidth]{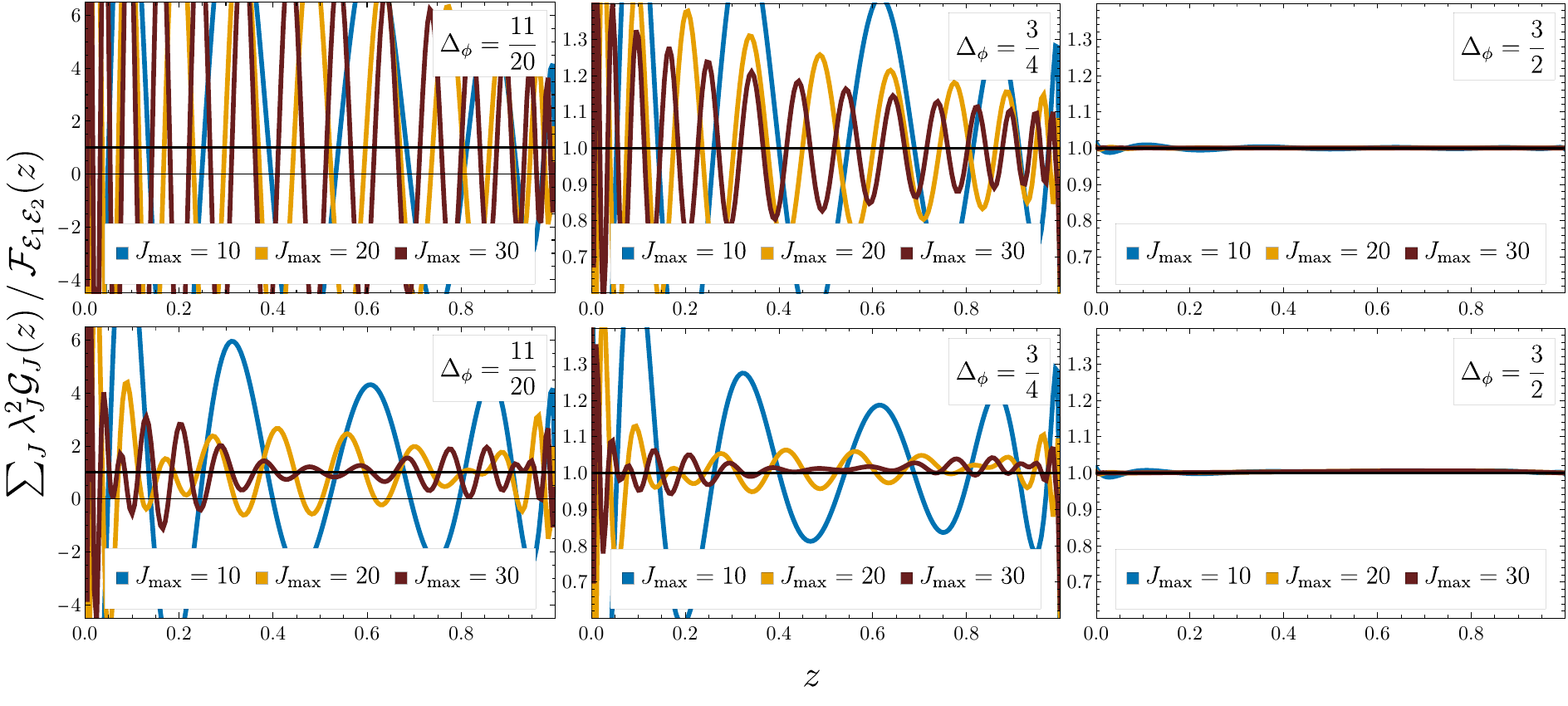}
\caption{\emph{Top:} Sum of energy correlator conformal blocks for double-twist operators $[\phi_1\phi_2]_{0,J}$ with $J\leq J_{\max}$ as a function of $z$ in a $d=3$ tensor product CFT, normalized by the exact correlator from~\eqref{eq:ProductCFTExactZ}, for different values of $J_{\max}$ and the source scaling dimension $\De_\phi$. \emph{Bottom:} Averaged version of the sum of conformal blocks, again normalized by the exact correlator. The result at each value of $z$ was obtained by averaging over the window $z \pm \fr{\ep}{2}$ with $\ep \propto z(1-z)$.}
\label{fig:ProductCFTConvergenceZ}
\end{figure}

This behavior can be seen explicitly in Figure~\ref{fig:ProductCFTConvergenceZ}, which shows the sum over double-twist conformal blocks in $d=3$ up to a maximum spin $J_{\max}$, normalized by the exact correlator~\eqref{eq:ProductCFTExactZ}, as a function of $z$. The top row shows the naive sum of conformal blocks for $\De_\phi = \fr{11}{20}$, which clearly does not converge, $\De_\phi = \fr{3}{4}$, which is right on the edge of convergence, and $\De_\phi=\fr{3}{2}$, which converges rather quickly. We can compare these results to the bottom row, which shows the same sums of conformal blocks, but with a moving average over the window $z \pm \half \ep \; z(1-z)$ with $\ep = 0.2$. All three sums now converge away from the endpoints, albeit more slowly for smaller values of $\De_\phi$.

In more general CFTs, we expect singular behavior in the limit $z \ra 0$ from the light-ray operator OPE~\cite{Hofman:2008ar,Kologlu:2019mfz}, which means that the sum over conformal blocks will generically need to be smeared against some test function to obtain a convergent function of $z$. This suggests that it may be difficult to naively  use data from the conformal bootstrap to construct the full $z$-dependent spectrum of the energy correlator. It would be interesting to study this convergence more carefully and determine if there is an optimal choice of basis for efficiently determining the full $z$-dependence of energy correlators in CFTs.


\section{Positivity Bounds on OPE Coefficients}
\label{sec:Bounds}

In this section we explore the use of the t-channel OPE to derive interesting bounds on CFT data from positivity of multi-point energy correlator observables. As discussed in the introduction, we will see that this provides a convenient way of repackaging the bounds from the ``interference effects" in the conformal collider \cite{Cordova:2017zej} as positivity of higher-point correlators. We also propose the use of energy correlators as positive measures to further strengthen these bounds.

The averaged null energy condition (ANEC) states that the expectation value of the energy operator in any state is non-negative,
\be
\<\psi|\Ecal(n)|\psi\> \geq 0 \textrm{ for all } |\psi\>.
\label{eq:ANEC}
\ee
This statement was proven for general unitary, Lorentz-invariant QFTs in~\cite{Faulkner:2016mzt} from monotonicity of relative entropy and in~\cite{Hartman:2016lgu} from causality. For general unitary CFTs, it was also proven in~\cite{Kologlu:2019bco} that energy operators inserted at different points on the celestial sphere commute,
\be
\comm{\Ecal(n_1)}{\Ecal(n_2)} = 0 \textrm{ for all } n_1 \neq n_2.
\ee

These two facts are sufficient to show that all higher-point energy correlators are also non-negative in any unitary CFT,\footnote{The ANEC~\eqref{eq:ANEC} is equivalent to the statement that the energy operator is a positive semi-definite operator (i.e.~its eigenvalues are all non-negative). If multiple insertions of $\Ecal(n)$ all commute, they can be simultaneously diagonalized, and the resulting composite operator $\Ecal(n_1) \cdots \Ecal(n_k)$ must also be positive semi-definite.}
\be
\<\psi|\Ecal(n_1) \cdots \Ecal(n_k)|\psi\> \geq 0 \textrm{ for all } |\psi\>.
\ee
In particular, the two-point function defined in~\eqref{eq:FDef} is non-negative for arbitrary $0\leq z \leq 1$,
\be
\Fcal_\Ecal(z) \geq 0.
\ee
The $SO(d-1)$ partial waves $\Chat^{(\fr{d-3}{2})}_\ell(x)$ in~\eqref{eq:FPartialWave} are normalized such that positivity of $\Fcal_\Ecal(z)$ guarantees the associated coefficients are bounded by the $\ell=0$ coefficient,
\be
|\Fcal_\Ecal^{(\ell)}| \leq \Fcal_\Ecal^{(0)},
\ee
which, as discussed in Sec.~\ref{sec:PartialWave}, is required by Ward identities to be $\Fcal_\Ecal^{(0)}=1$. Furthermore, unitarity ensures that all partial wave coefficients are non-negative, such that we obtain the general bounds \cite{Fox:1978vw} (see also \cite{sasha_positive})
\be
0 \leq \Fcal_\Ecal^{(\ell)} \leq 1 \textrm{ for all } \ell.
\label{eq:PartialWaveBounds}
\ee

If we now look at the conformal block decomposition of these partial wave coefficients, we find that positivity of energy correlators leads to an infinite set of bounds on the OPE coefficients of any unitary CFT,\footnote{For simplicity, here we are specifically considering CFTs in $d\geq 4$, where there is only one three-point function tensor structure for each exchanged operator $\Ocal$, given by~\eqref{eq:ThreePtL}. Similar bounds can be derived for theories in $d=3$, as we'll see for the specific example of the 3d Ising CFT below.}
\be
\boxed{\sum_\Ocal \lambda_{\phi T\Ocal}^2 \; \Gcal_\Ocal^{(\ell)} \leq 1,}
\label{eq:SumRules}
\ee
where the sum is over all primary operators $\Ocal$ appearing in the $\phi \times T_{\mu\nu}$ OPE, with all conformal block coefficients $\Gcal_\Ocal^{(\ell)} \geq 0$. This bound is one of the primary results of this paper.

It will be interesting to explore the full consequences of our bound in future work~\cite{UsMoments}, but here we will instead discuss the simpler, weaker bounds obtained by only focusing on the contribution of operators with spin $J=\ell$,
\be
\sum_{\textrm{spin-}J \, \Ocal} \lambda_{\phi T\Ocal}^2 \; \Gcal_\Ocal^{(J)} \leq 1,
\label{eq:SpinJBound}
\ee
where now the sum is over primary operators of spin $J$, with the simpler conformal block coefficients,
\be
\Gcal_\Ocal^{(J)} = \fr{\pi^d \; J! \; (d-1)^2 \; \G^2(\fr{d-2}{2}) \; (d-1)_J^2 \; \G(\De_\phi) \; \G(\De_\phi-\fr{d-2}{2}) \; \G(\De+J) \; \G(\De-\fr{d-2}{2}) \; (\De-1)_J}{2^{3J} \De^2(\De-1)^2 \; (\fr{d-1}{2})_J \; \G^2(\fr{\De_\phi-\tau+d+2}{2}) \; \G^2(\fr{\De+J-\De_\phi+d-2}{2}) \; \G^2(\fr{\De+J+\De_\phi-2}{2}) \; \G^2(\fr{\De+J+\De_\phi-d+2}{2})}.
\ee

As a concrete example, let's consider the contribution of the stress tensor, with spin $J=2$. The bound~\eqref{eq:SpinJBound} leads to the constraint
\be
\boxed{\lambda_{\phi TT}^2 \leq \fr{16 \; \G^4(\fr{\De_\phi}{2}+2) \; \G^2(\fr{d+\De_\phi}{2}) \; \G^2(d-\fr{\De_\phi}{2})}{d^2(d-1)^2 \; \pi^d \; \G(\fr{d}{2}) \; \G^2(\fr{d-2}{2}) \; \G(d+1) \; \G(\De_\phi) \; \G(\De_\phi-\fr{d-2}{2})} \; c_T,}
\label{eq:OTTBound}
\ee
where the factor of $c_T$ comes from our normalization convention for the stress tensor, see~\eqref{eq:TNorm}. We can compare this bound to the similar one obtained in~\cite{Cordova:2017zej},
\be
\lambda_{\phi TT}^2 \leq \fr{(d-2)^2 \; \G^4(\fr{\De_\phi}{2}+2) \; \G^2(\fr{d+\De_\phi}{2}) \; \G^2(d-\fr{\De_\phi}{2})}{d(d-1)^3 \; \pi^{2d} \; \G(\fr{d}{2}) \; \G(d+1) \; \G(\De_\phi) \; \G(\De_\phi-\fr{d-2}{2})} \; n_B,
\label{eq:InterferenceBound}
\ee
where $n_B$, $n_F$, and $n_V$ are the independent OPE coefficients for the stress tensor three-point function in $d\geq 4$, which are related to $c_T$ by~\cite{Osborn:1993cr}
\be
c_T = \fr{\G^2(\fr{d}{2})}{4\pi^d} \Big( \fr{d}{d-1} n_B + \fr{d}{2} n_F + \fr{d^2}{2} n_V \Big).
\ee
The ``interference effects'' bound~\eqref{eq:InterferenceBound} from~\cite{Cordova:2017zej} was derived by imposing the ANEC in every state created by a linear combination of $\phi$ and $T_{\mu\nu}$. This bound is strictly stronger than our result~\eqref{eq:OTTBound}, though the two bounds coincide for theories with $n_F = n_V = 0$.

The reason our positivity bound~\eqref{eq:OTTBound} is weaker than that obtained in~\cite{Cordova:2017zej} is that the energy two-point function in the background of $\phi$ only probes one particular linear combination of the components of $T_{\mu\nu}$. If we instead considered the energy two-point function in the background created by an arbitrary polarization of $T_{\mu\nu}$ and focused on the positivity constraints for scalar operator conformal blocks in the $T \times T$ OPE, we would recover the stronger bound~\eqref{eq:InterferenceBound}.

We therefore see that the conformal block decomposition of higher-point energy correlators provides an alternative means of deriving interference effect bounds, without the need to consider external states created by linear combinations of multiple operators. It would be interesting to further study these bounds for external states with spin or for correlators involving additional insertions of $\Ecal(n)$.

As a final demonstration of these positivity bounds, let's consider the specific case of the 3d Ising CFT. As discussed in Sec.~\ref{sec:Blocks}, in $d=3$ there are two allowed three-point function tensor structures, which are respectively even and odd under parity. Because this particular CFT is parity-invariant, all operators have a well-defined charge $\pm 1$ under parity transformations. The positivity bounds~\eqref{eq:SumRules} can therefore be divided into separate sums over parity-even and parity-odd intermediate operators,
\be
\sum_{\Ocal^+} \lambda_{\phi T\Ocal^+}^2 \; \Gcal_{\Ocal^+}^{(\ell^+)} + \sum_{\Ocal^-} \lambda_{\phi T\Ocal^-}^2 \; \Gcal_{\Ocal^-}^{(\ell^-)} \leq 1,
\label{eq:3dSumRules}
\ee
where we've assumed the external source $\phi$ is parity-even.\footnote{More generally, the first sum is over operators with the same parity charge as $\phi$ and the second sum over those with the opposite charge.}

\begin{figure}[t!]
\centering
\includegraphics[width=.95\linewidth]{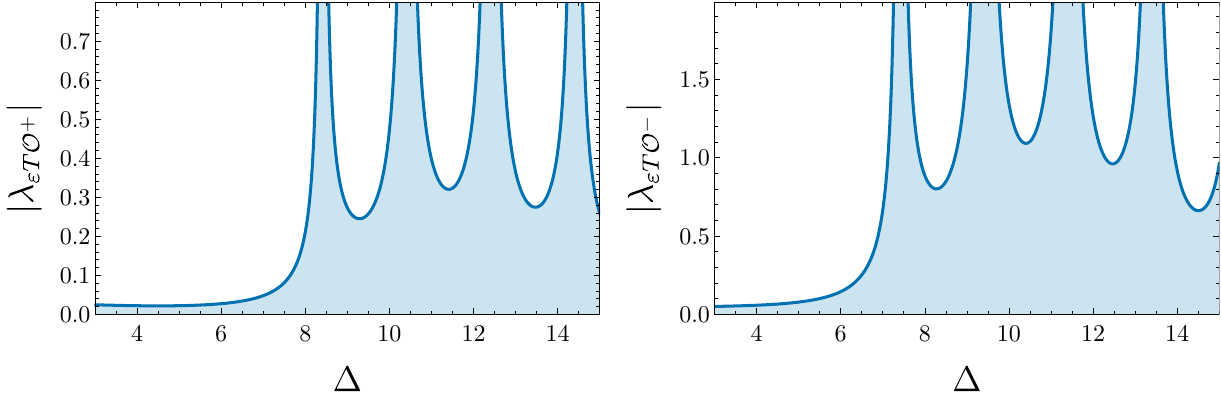}
\caption{Positivity bounds on $|\lambda_{\vep T\Ocal}|$ for any parity-even (\emph{left}) or parity-odd (\emph{right}) primary operator $\Ocal$ with spin $J=2$ in the 3d Ising CFT (other than the stress tensor $T_{\mu\nu}$) as a function of its scaling dimension $\De$. The shaded region indicates the allowed OPE coefficient magnitudes, using the numerical bootstrap data~\eqref{eq:BootstrapData} obtained in~\cite{Chang:2024whx}. Similar bounds can be derived for operators with spin $J\geq 3$.}
\label{fig:IsingBounds}
\end{figure}

The most precise results for the 3d Ising CFT were obtained in~\cite{Chang:2024whx} by applying the numerical conformal bootstrap to all four-point functions built from $\s$, $\vep$, and $T_{\mu\nu}$, yielding the following data relevant for our purposes:
\be
\De_\vep = 1.41262528(29), \quad c_T=0.00899103927(40), \quad \lambda_{\vep TT} = 0.0085712937(38).
\label{eq:BootstrapData}
\ee
Because $\vep$ is parity-even, we can plug in these values for $\De_\vep$ and $c_T$ to~\eqref{eq:OTTBound} to obtain the resulting bound\footnote{Our normalization convention for $\lambda_{\vep TT}$ differs from that of~\cite{Chang:2024whx}, which uses $\lambda_{\vep \hat{T}\hat{T}} \equiv \fr{1}{c_T} \lambda_{\vep TT}$. The equivalent bound in those conventions is $|\lambda_{\vep \hat{T}\hat{T}}| \leq 0.988$.}
\be
\boxed{|\lambda_{\vep TT}| \leq 0.00888.}
\ee
For comparison, the equivalent bound obtained from~\cite{Cordova:2017zej} is $|\lambda_{\vep TT}| \leq 0.00882$. As we can see from the bootstrap result~\eqref{eq:BootstrapData}, $\lambda_{\vep TT}$ almost saturates the bound, which means that $F_\Ecal^{(2)}$ in the Ising CFT must be very close to $1$,
\be
\Fcal_\Ecal^{(2)} \geq \lambda_{\vep TT}^2 \; \Gcal_T^{(2)} = 0.932.
\ee
We can in turn plug the bootstrap value for $\lambda_{\vep TT}$ into~\eqref{eq:3dSumRules} to constrain all other spin-two operators appearing in the $\vep \times T_{\mu\nu}$ OPE,
\be
\boxed{
\ba
&\sum_{\textrm{spin-}2 \, \Ocal^+ \neq T} \fr{9\pi^5 (\De+1) \; \G(2\De_\vep-1) \; \G(2\De-2)}{2^{2\De_\vep+2\De-3} \G^2(\fr{\De_\vep-\De+7}{2}) \; \G^2(\fr{\De-\De_\vep+3}{2}) \; \G^2(\fr{\De+\De_\vep}{2}) \; \G^2(\fr{\De+\De_\vep+1}{2})} \; \lambda_{\vep T\Ocal^+}^2 \\
& \quad + \sum_{\textrm{spin-}2 \, \Ocal^-} \fr{9\pi^5 (\De+1) \; \G(2\De_\vep-1) \; \G(2\De-2)}{2^{2\De_\vep+2\De-1} \G^2(\fr{\De_\vep-\De+6}{2}) \; \G^2(\fr{\De-\De_\vep+4}{2}) \; \G^2(\fr{\De+\De_\vep}{2}) \; \G^2(\fr{\De+\De_\vep+1}{2})} \; \lambda_{\vep T\Ocal^-}^2 \leq 0.068.
\ea
}
\label{eq:IsingBounds}
\ee

\begin{figure}[t!]
\centering
\begin{tabular}{c}
\includegraphics[width=.9\linewidth]{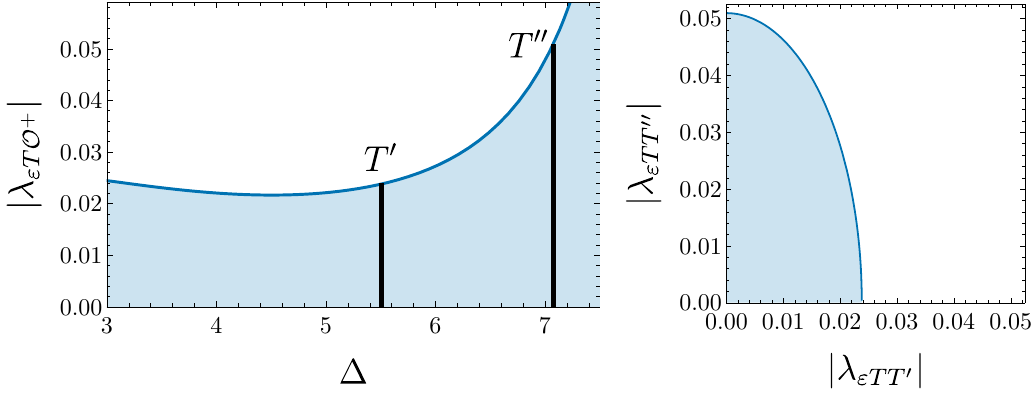} \\
\includegraphics[width=.53\linewidth]{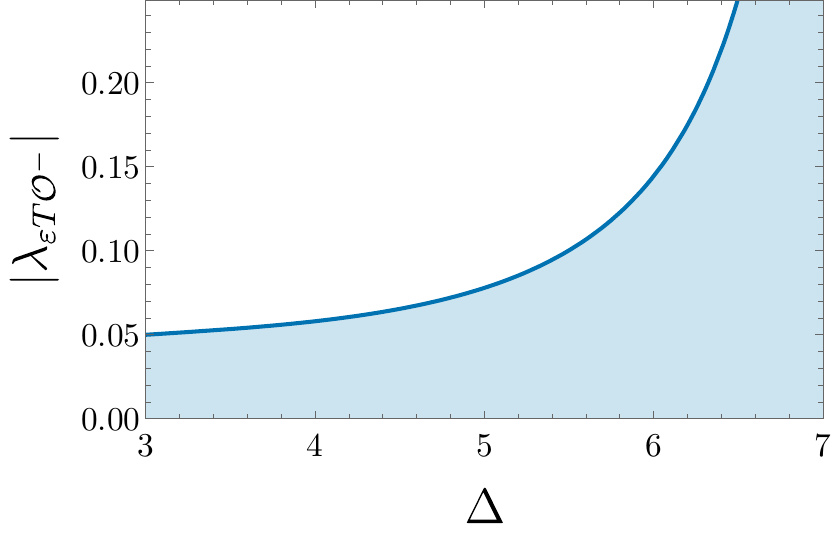}
\end{tabular}
\caption{Zoomed-in version of the positivity bounds from Figure~\ref{fig:IsingBounds} for parity-even (\emph{top left}) and parity-odd (\emph{bottom}) operators, with vertical black lines highlighting the resulting bounds at values of $\De$ for the known parity-even spin $J=2$ operators $T'$ and $T''$ from~\eqref{eq:BootstrapData2}. \emph{Top right:} Joint bounds on the OPE coefficients for $T'$ and $T''$.}
\label{fig:IsingBoundsZoom}
\end{figure}

Figure~\ref{fig:IsingBounds} shows the resulting bounds on individual OPE coefficients, where the shaded region indicates the values allowed by~\eqref{eq:IsingBounds}. As we can see, the bound weakens significantly near particular values of $\De$ where the energy correlator conformal blocks vanish, corresponding to the $[\vep T]_{n,2}$ double-twist scaling dimensions $\De_\vep + 7 + 2n$ for parity-even operators and $\De_\vep + 6 + 2n$ for parity-odd ones~\cite{Kologlu:2019bco}.\footnote{One would naively expect the parity-even conformal blocks to also vanish at $\De_\vep+3$ and $\De_\vep+5$ (or $\De_\vep+4$ in the parity-odd case), but these zeroes have been removed by our choice of normalization of the OPE coefficients $\lambda_{\vep T\Ocal}^{(\pm)}$ in~\eqref{eq:ThreePtPos} and \eqref{eq:ThreePtPosOdd}, corresponding to the factor of~$(\tau-\De_\phi-d)(\tau-\De_\phi-d+2)$ in the denominator in~\eqref{eq:WardOPE} and $\tau-\De_\phi-2$ in~\eqref{eq:WardOPEOdd}. We thank Rajeev Erramilli for useful discussions on this point.}

Figure~\ref{fig:IsingBoundsZoom} shows a zoomed-in version focusing on low-dimension operators, where the bounds are strongest. The values of $\De$ corresponding to the known spin-two primary operators $T'$ and $T''$ in the 3d Ising CFT are indicated with vertical black lines, using the scaling dimensions~\cite{Simmons-Duffin:2016wlq}
\be
\De_{T'} = 5.50915(44), \quad \De_{T''}=7.0758(58).
\label{eq:BootstrapData2}
\ee
However, these two operators cannot both saturate their respective bounds simultaneously, but instead must satisfy a combined bound from~\eqref{eq:IsingBounds}. The resulting allowed region for these two OPE coefficients is shown in the upper right plot of Figure~\ref{fig:IsingBoundsZoom}. As more spin-$J \geq 2$ OPE coefficients are computed, these positivity bounds will continue to strengthen, further reducing the allowed values for all remaining OPE coefficients.

It is worth emphasizing that the results in Figures~\ref{fig:IsingBounds} and \ref{fig:IsingBoundsZoom} can all be computed almost instantaneously on a laptop, given the initial data~\eqref{eq:BootstrapData} and \eqref{eq:BootstrapData2}. While the bounds are not as precise as those typically obtained with numerical bootstrap methods (e.g.~\cite{Chang:2024whx}), it is intriguing that energy correlators provide a very efficient means of rigorously constraining stress tensor OPE coefficients. It would be exciting to try to combine the two methods in the future, using positivity as a guide as to the allowed range of OPE coefficients for the numerical bootstrap.


\subsection{Energy Correlators as Positive Measures}

In the previous section, we have exploited the simplest possible manifestation of the positivity of the two-point energy correlator to derive bounds on sums of OPE coefficients. Here we briefly discuss how this can be generalized by viewing the energy correlator as a positive measure,\footnote{We thank Ryan Lanzetta for motivating discussions about positive measures.} leaving exploration of the implied bounds for concrete theories to future work~\cite{UsMoments}.

The general positivity bounds~\eqref{eq:PartialWaveBounds} on partial wave coefficients $F_\Ecal^{(\ell)}$ are overly conservative, as a generic combination of $F_\Ecal^{(\ell)}$ satisfying~\eqref{eq:PartialWaveBounds} will not correspond to a positive function of $z$. Instead one can derive more stringent bounds on the relative sizes of coefficients with different $\ell$ by viewing them as moments of a positive measure constructed from the energy correlator
\be
d\mu(z)=\Fcal_\Ecal(z) \, dz.
\ee
Similar problems have been explored extensively in the context of the S-matrix bootstrap for bounding effective field theory coefficients \cite{Bellazzini:2020cot,Arkani-Hamed:2020blm,Caron-Huot:2020cmc}, and those methods can be straightforwardly adapted to energy correlators to further improve the bounds presented in the previous section.

Interestingly, the resulting ``moment bounds'' relate infinite sums of OPE coefficients at different spins. As emphasized above, these positivity constraints are a priori distinct from crossing, being purely based on positivity of energy flux. It will be exciting to explore the numerical implementation of such bounds, or develop other approaches to their study, such as developing an analog of extremal functionals \cite{Mazac:2016qev,Mazac:2019shk,Mazac:2018mdx,Caron-Huot:2020adz}. In the particular case of the 3d Ising CFT, it will be interesting to sharpen the bounds derived in the previous section and compare with data from the conformal bootstrap.


\section{Discussion}
\label{sec:Discussion}

In this work, we have explored the t-channel OPE of energy correlators. We computed all two-point energy correlator conformal blocks associated with the exchange of traceless symmetric operators $\Ocal$ in the $\phi \times \Ecal$ OPE, including both parity-even and parity-odd contributions.  Our result extends the calculation of the scalar event shape blocks in the literature, allowing much broader applications, in particular to theories where the stress tensor is not related to a scalar operator by supersymmetry.  In the context of a tensor product CFT, we performed a simple study of the convergence of the OPE, both at the level of the partial wave coefficients, as well as for the full distribution of the energy correlator as a function of the physical angle between detectors. We found that the convergence for the angular distribution converges poorly, suggesting that more sophisticated methods are required. However, for the case of the individual partial waves, we observed rapid convergence. This motivates exploring the full extent to which coarse-grained information on the angular dependence of energy correlators can be approximated with low-dimension conformal blocks and their calculation in specific CFTs using data for the scaling dimensions and OPE coefficients of local operators.

We also showed that the positivity of higher-point energy correlators, combined with the t-channel OPE, provides a simple way of packing an infinite set of ``interference effects" in the conformal collider setup, and we used this to derive bounds on sums of OPE coefficients in generic CFTs. We explored this in the particular case of the 3d Ising CFT, deriving new bounds on both parity-even and parity-odd OPE coefficients. Even for the particular case of the two-point correlator considered here, we have only explored the minimal consequences of the positivity bounds presented in Sec.~\ref{sec:Bounds}. We highlighted the ability to use the energy correlator distribution as a positive measure, similar to what has been done in the context of the S-matrix bootstrap (see e.g.~\cite{Bellazzini:2020cot,Caron-Huot:2020cmc,Arkani-Hamed:2020blm}). It will be very exciting to determine the full space of OPE coefficients allowed by energy positivity, both in general CFTs and in individual theories where some information on the spectrum is known. 

There are a number of directions in which the studies of this paper could be extended:

{\bf{Further Exploring Bounds on CFT Data:}} In this paper we only considered the concrete application of our bounds to the case of the 3d Ising CFT. It would be interesting to extend this to a variety of other CFTs for which there exist OPE data from the conformal bootstrap or fuzzy sphere regularization, for example the $O(N)$~\cite{Kos:2015mba,Kos:2016ysd,Chester:2019ifh,Guo:2025odn} or Gross-Neveu-Yukawa~\cite{Erramilli:2022kgp} models. Chern-Simons matter theories~\cite{Zhou:2025rmv} were also explored in \cite{Cordova:2017zej}, and it would be interesting to consider them using our approach. More ambitiously, our bounds are particularly well suited for applications at finite $N$, as positivity constrains the full non-perturbative OPE coefficients, rather than individual terms in an expansion in $1/N$. It would be interesting to consider the resulting bounds in non-planar ABJM~\cite{Chester:2014fya,Agmon:2017xes,Agmon:2019imm,Alday:2021ymb} and $\Ncal=4$ SYM~\cite{Beem:2013qxa,Beem:2016wfs,Chester:2021aun,N4_bootstrap}, or for finite $N_c$ Banks-Zaks or finite $N_f$ QED$_3$~\cite{Chester:2016wrc,He:2021sto,Albayrak:2021xtd}.

{\bf{Spinning Sources:}} A straightforward extension of the present work is the calculation of conformal blocks for external sources with nonzero spin, in order to generalize the positivity bounds of Sec.~\ref{sec:Bounds} to all possible OPE coefficients involving the stress-energy tensor $T_{\mu\nu}$. This would effectively generalize our approach to a ``mixed correlator" bootstrap. There are a number of interesting targets when the source is a conserved vector current, or R-current, $J_\mu$, including mixed anomalies \cite{Erdmenger:1999xx,Cordova:2017zej} and transport coefficients \cite{Son:2009tf,Gooth:2017mbd}. While in principle the calculation of the relevant blocks can be done following the same steps used here (Fourier transforming the allowed position space tensor structures to momentum space), a more efficient and systematic approach would be to use ``weight shifting'' operators~\cite{Karateev:2017jgd,Baumann:2019oyu,Bzowski:2022rlz,Caloro:2022zuy} to directly convert a momentum space three-point function with two scalars $\<\phi\;\Ecal\;\phi'\>$ into a correlator involving two arbitrary-spin operators $\<\Ocal\;\Ecal\;\Ocal'\>$, in order to obtain the resulting conformal blocks.\footnote{We thank Denis Karateev for useful discussions on this point.} While most of the applications of weight shifting operators for the conformal bootstrap have been in position space, the momentum space weight shifting operators developed in the study of cosmological correlators \cite{Arkani-Hamed:2018kmz,Baumann:2019oyu,Baumann:2020dch} seem particularly useful for systematizing the study of spinning t-channel conformal blocks for energy correlators.

A further advantage of generalizing to spinning external states is that one can use the resulting conformal blocks to analyze the ``superconvergence'' sum rules which follow from commutativity of energy operators~\cite{Kologlu:2019bco},
\be
\<\comm{\Ecal(n_1)}{\Ecal(n_2)}\> = 0.
\ee
By decomposing the two differently ordered correlators in this expression as sums over conformal blocks, one obtains a nontrivial constraint on the associated OPE coefficients, the consequences of which would be interesting to study in concrete examples like the 3d Ising CFT. For a scalar external source, the contribution of each conformal block to this commutator vanishes trivially as their only dependence on the polarization vectors is through the cross-ratio $z$, which is invariant under the exchange $n_1 \lra n_2$. However, for external operators with spin, one can construct additional cross-ratios involving the external polarization vector, such that each individual conformal block gives a nonzero contribution to the commutator, which must therefore all cancel in the full sum over intermediate states.

{\bf{Higher Spin Detectors:}}
Another promising direction is the extension of these conformal block calculations to more general detectors, either built from integrating the stress tensor against some kernel or from other local operators, such as conserved currents. For example, it was shown in~\cite{Hartman:2016lgu} that the detector built from the lowest-dimension operator $\Ocal_{\min}$ for every even spin $J \geq 2$ is positive semi-definite,
\be
\<\psi|\Dcal_{\Ocal_{\min}}(n)|\psi\> \geq 0.
\ee
In other words, the ANEC is only one example of an infinite set of positivity conditions satisfied in general CFTs. ``Higher spin ANEC operators'' of spins $J_1$ and $J_2$ commute when inserted at different points on the celestial sphere, assuming that $J_1+J_2>J_0+1$, where $J_0$ is the Regge intercept~\cite{Kologlu:2019bco}. Their higher-point functions are therefore also positive, allowing us to derive straightforward generalizations of our positivity constraints. This motivates further exploration of Regge trajectories in different theories, and their relation with detector operators \cite{Ekhammar:2024neh,Klabbers:2023zdz,Brizio:2024nso,Gromov:2015wca,Gromov:2015vua,Alfimov:2014bwa,Ekhammar:2025vig,Caron-Huot:2022eqs,Li:2025knf}.

{\bf{Higher Points:}}
An alternative route to access general OPE coefficients of the stress tensor is to consider higher-point energy correlators in the background of a scalar source. For example, the three-point energy correlator can be written as the following schematic double-sum over intermediate operators,
\be
\fr{\<\Ecal(n_1) \; \Ecal(n_2) \; \Ecal(n_3)\>}{\<\Ecal(n_1)\>\<\Ecal(n_2)\>\<\Ecal(n_3)\>} = \sum_{\Ocal,\Ocal'} \lambda_{\phi T\Ocal} \; \lambda_{\phi T\Ocal'} \; \lambda_{\Ocal T \Ocal'} \; \Gcal_{\Ocal\Ocal'}(z_i),
\ee
where $z_i$ are the various dimensionless cross-ratios built from the external momentum and polarization vectors. The associated conformal blocks can similarly be built from momentum space three-point functions of local operators, following the procedure used in this work. As this three-point energy correlator must be non-negative for all choices of $n_i$, one obtains new positivity constraints involving more general OPE coefficients than those accessible with the two-point correlator. Furthermore, these new constraints are sensitive to the relative signs of OPE coefficients, not just the absolute values.

{\bf{Mixed Channel Bootstrap:}} One long-term motivation underlying this work is the construction of an ``energy correlator conformal bootstrap'' by equating the t-channel conformal block expansion we consider here to the s-channel expansion in terms of celestial blocks obtained from the light-ray OPE~\cite{Kologlu:2019mfz}. As highlighted in Sec.~\ref{sec:Intro}, these two expansions depend on very different aspects of the CFT data, with one OPE depending on the spectrum of light-ray operators, and the other on the spectrum of local operators. In particular the light-ray OPE depends on the analytic continuation of OPE coefficients along Regge trajectories to unphysical spins, and it would be very interesting to explore the constraints arising from the equality of these two different expansions. While the complete formulation of such a bootstrap seems challenging, a truncated bootstrap \cite{Gliozzi:2013ysa,Gliozzi:2014jsa,Li:2017ukc} similar to that considered in the case of five-point correlators may be feasible \cite{Poland:2023vpn,Poland:2025ide}.

{\bf{Energy Correlators as Positive Measures:}} In this paper we only numerically studied the simplest possible bound arising from positivity of the energy correlator. We highlighted that more generally energy correlators can be viewed as positive measures, allowing more stringent constraints to be derived on OPE data~\cite{UsMoments}. It will interesting to explore these numerically. This would be particularly interesting in the context of the 3d Ising CFT, and given the conformal blocks derived in this paper, is now possible. This approach also naturally generalizes to higher-point correlators. Additionally, it would be interesting to explore other approaches to deriving constraints from posivity of the energy correlators, for example extremal functionals \cite{Caron-Huot:2020adz}.

{\bf{Constraints on dS/AdS Effective Actions:}} In this paper we have focused on the CFT interpretation of our bounds. Bounds on OPE coefficients involving stress tensors are particularly interesting for sharpening the understanding of the emergence of bulk locality in AdS/CFT \cite{Heemskerk:2009pn}. While the $\<TTT\>$ three-point function has been extensively studied \cite{Heemskerk:2009pn,Costa:2017twz,Kulaxizi:2017ixa,Afkhami-Jeddi:2016ntf,Camanho:2014apa,Hofman:2008ar,Buchel:2009sk}, constraints on $\<\mathcal{O}TT\>$, with $\mathcal{O}$ a scalar or non-conserved spinning operator, have received less attention \cite{Meltzer:2017rtf,Cordova:2017zej}. Our bounds from higher-point energy correlators seem ideally suited for their study. Ref.~\cite{Cordova:2017zej} also explored their bounds in the context of effective actions in dS, to derive bounds on higher derivative corrections \cite{Maldacena:2011nz,Soda:2011am,Bartolo:2017szm} to non-gaussianities \cite{Maldacena:2002vr}. It would be interesting to further explore the implications of our bounds, or more general implications of higher-point positivity for theories in (anti-)de Sitter.

{\bf{Extension to Gravity:}} There has been tremendous efforts to constrain higher derivative couplings in gravity, using the S-matrix bootstrap \cite{Camanho:2014apa,Guerrieri:2021ivu,Guerrieri:2022sod,Caron-Huot:2021rmr}. In the case of $d=4$, this is made more challenging due to the presence of infrared divergences. An appealing feature of energy correlator observables is that they are well defined in theories that do not admit correlation functions of local operators or S-matrices.  Energy correlators in quantum gravity have recently been explored, and explicitly computed in \cite{Gonzo:2020xza,Gonzo:2023cnv,Herrmann:2024yai}. It will be particularly interesting to understand how positivity can be exploited in the gravitational case.


\section*{Acknowledgments}    

It is a pleasure to thank Rajeev Erramilli, Kara Farnsworth, Liam Fitzpatrick, Denis Karateev, Ami Katz, Murat Kologlu, Ryan Lanzetta, Joseph Lap, Miguel Paulos, David Poland, David Simmons-Duffin, Hidde Stoffels, Petar Tadi\'{c}, Zahra Zahraee, and Sasha Zhiboedov for valuable discussions. BM is supported by the US Department of Energy through the Los Alamos National Laboratory and by LANL’s Laboratory Directed Research and Development (LDRD) program. Los Alamos National Laboratory is operated by Triad National Security, LLC, for the National Nuclear Security Administration of the U.S. Department of Energy (Contract No.~89233218CNA000001). IM is supported by the DOE Early Career Award DE-SC0025581, and the Sloan Foundation. MW is supported by the Royal Society University Research Fellowship URF{\textbackslash}R1{\textbackslash}221905 and is grateful to \'{E}cole Normale Sup\'{e}rieure for hospitality while this work was completed.
We would also like to thank the Mainz Institute for Theoretical Physics and the Simons Center for Geometry and Physics for hospitality. This work was performed in part at the Kavli Institute for Theoretical Physics, which is supported by the National Science Foundation grant PHY-2309135.


\appendix

\section{Details of Conformal Block Calculations}
\label{app:Details}

Here we provide more details on the computation of energy correlator conformal blocks presented in Sec.~\ref{sec:Method} and \ref{sec:Blocks}. This calculation is based on the method developed in~\cite{Kologlu:2019bco} for constructing conformal blocks for general light-ray operators.


\subsection{Review of Scalar Detector Conformal Blocks}
\label{sec:ScalarBlocks}

As a useful warmup, let's first review the simpler calculation of conformal blocks for scalar detectors, which was initially derived in~\cite{Kologlu:2019bco}. This will introduce all the key steps for computing energy correlator conformal blocks, but with somewhat less complicated expressions.

Given a scalar primary operator $D$ with dimension $\De_D$, we can define the associated detector operator~\cite{Belitsky:2013xxa},\footnote{This detector is related to the light transform of $D$ defined in~\cite{Kravchuk:2018htv} by an overall factor of 2,
\be
\mathbf{L}[D](\infty;n) \equiv \lim_{x\ra\infty} \int_{-\infty}^\infty d\alpha \; (-\alpha)^{-\De_D} \; D\Big(x-\fr{n}{\alpha}\Big) = 2\Dcal(n).
\ee}
\be
\Dcal(n) = \int_{-\infty}^\infty \! d(x\cdot n) \lim_{x\cdot\bar{n}\ra\infty} \! \bigg(\fr{x\cdot\bar{n}}{n\cdot\bar{n}}\bigg)^{\De_D} \; D(x),
\ee
where the auxiliary null vector $\bar{n}^\mu$ used to move the detector to infinity is arbitrary (so long as $n\cdot\bar{n}\neq0$) and does not appear in correlators of $\Dcal(n)$.

We can then define the two-point function of this detector in the background created by another scalar primary operator $\phi$,
\be
\Fcal_\Dcal(z) \equiv \fr{\<\phi(p)|\Dcal(n_1) \; \Dcal(n_2)|\phi(p)\>}{\<\phi(p)|\phi(p)\> \; \<\Dcal(n_1)\> \; \<\Dcal(n_2)\>} = \sum_\ell \Fcal_\Dcal^{(\ell)} \; \Chat_\ell^{(\fr{d-3}{2})}(1-2z),
\ee
where $z$ is the dimensionless cross-ratio defined in~\eqref{eq:CrossRatio} and the one-point function is fixed by conformal invariance (up to an overall OPE coefficient),
\be
\<\Dcal(n)\> = \fr{\lambda_{\phi D\phi} \; \pi \; \G(\De_D-1) \; \G(\De_\phi) \; \G(\De_\phi-\fr{d-2}{2})}{2^{\De_D-2} \G^2(\fr{\De_D}{2}) \; \G(\De_\phi-\fr{\De_D}{2}) \; \G(\De_\phi+\fr{\De_D-d}{2})} \fr{p^{\De_D-2}}{(p\cdot n)^{\De_D-1}}.
\ee

This two-point function can be decomposed in terms of conformal blocks in the ``source-detector'' channel by inserting a complete set of intermediate states,
\be
\ba
\Fcal_\Dcal(z) &= \sum_\Ocal \lambda_{\phi D\Ocal}^2 \; \Gcal_\Ocal(z) \\
&= \sum_\Ocal \int\fr{d^dp'}{(2\pi)^d} \fr{\<\phi(p)|\Dcal(n_1)|\Ocal_{\mu_1\cdots\mu_J}(p')\>\;\Pi_\Ocal^{\mu_1\cdots\mu_J;\nu_1\cdots\nu_J}(p')\;\<\Ocal_{\nu_1\cdots\nu_J}(p')|\Dcal(n_2)|\phi(p)\>}{\<\phi(p)|\phi(p)\> \; \<\Dcal(n_1)\> \; \<\Dcal(n_2)\>}.
\ea
\label{eq:ScalarDecomp}
\ee
The projector $\Pi_\Ocal(p)$ can roughly be thought of as the inverse of the two-point function of the intermediate operator,
\be
\Pi_\Ocal^{\mu_1\cdots\mu_J;\nu_1\cdots\nu_J}(p) \sim \fr{1}{\<\Ocal_{\mu_1\cdots\mu_J}(p)|\Ocal_{\nu_1\cdots\nu_J}(p)\>}.
\ee
We can make this statement more precise by decomposing $\Pi_\Ocal(p)$ in terms of projectors onto the spin-$\ell$ representations of $SO(d-1)$,\footnote{The $SO(d-1)$ projectors can be rewritten in index-free notation as
\be
\Pi_\ell^J(p;n_1,n_2) \equiv (-1)^\ell \fr{(p\cdot n_1)^J(p\cdot n_2)^J}{p^{2J}} \; \Pi_\ell^J(z),
\ee
with $\Pi_\ell^J(z)$ defined in~\eqref{eq:PiJLdef}. As a concrete example, the projectors for $J=1$ are
\be
\Pi_0^{\mu;\nu}(p) = \fr{p^\mu p^\nu}{p^2}, \qquad \Pi_1^{\mu;\nu}(p) = \eta_{\mu\nu}-\fr{p^\mu p^\nu}{p^2}.
\ee
}
\be
\Pi_\Ocal^{\mu_1\cdots\mu_J;\nu_1\cdots\nu_J}(p) = \fr{1}{p^{2\De-d}} \sum_{\ell=0}^J \fr{(-1)^\ell}{F_\Ocal^{(\ell)}} \; \Pi_\ell^{\mu_1\cdots\mu_J;\nu_1\cdots\nu_J}(p),
\ee
with coefficients given by the inverse of the two-point function coefficients $F_\Ocal^{(\ell)}$ from~\eqref{eq:TwoPtCoeffs}.

If we also decompose the three-point functions in terms of these projection operators,
\be
\ba
\<\phi(p)|\Dcal(n_1)|\Ocal_{\mu_1\cdots\mu_J}(p')\> &\equiv (2\pi)^d\de^d(p-p') \; \Theta(p_0) \; \Theta(p^2) \; \lambda_{\phi D\Ocal} \; \<\Dcal(n_1)\> \; \fr{p^{\De+J+\De_\phi-d}}{(p\cdot n_1)^J} \\
&\qquad \times \sum_{\ell=0}^J (-1)^\ell \; \Ccal_{\phi\Dcal\Ocal}^{(\ell)} \; n_{1\alpha_1} \cdots n_{1\alpha_J} \Pi^{\alpha_1\cdots\alpha_{J}}_{\ell\phantom{\cdots\alpha_J}\mu_1\cdots\mu_{J}}(p),
\ea
\ee
we can use the identity
\be
\Pi^{\mu_1\cdots\mu_{J}}_{\ell\phantom{\cdots\mu_J}\alpha_1\cdots\alpha_{J}}(p) \; \Pi_{\ell'}^{\alpha_1\cdots\alpha_{J};\nu_1\cdots\nu_{J}}\!(p) = \de_{\ell\ell'} \; \Pi_\ell^{\mu_1\cdots\mu_{J};\nu_1\cdots\nu_{J}}(p),
\ee
to rewrite the conformal blocks in~\eqref{eq:ScalarDecomp} as
\be
\Gcal_\Ocal(z) = \fr{p^{2J}}{(p\cdot n_1)^J (p\cdot n_2)^J} \sum_{\ell=0}^J (-1)^\ell \; \fr{|\Ccal_{\phi\Dcal\Ocal}^{(\ell)}|^2}{F_\phi \; F_\Ocal^{(\ell)}} \; n_{1\mu_1} \cdots n_{1\mu_J} n_{2\nu_1} \cdots n_{2\nu_J} \; \Pi_\ell^{\mu_1\cdots\mu_{J};\nu_1\cdots\nu_{J}}(p).
\label{eq:ScalarBlock}
\ee

We therefore need to compute the scalar detector three-point function
\be
\Ccal_{\phi\Dcal\Ocal}(z) = \int d^dx_0 \; e^{ip\cdot x_0} \int_{-\infty}^\infty \! d(x_1\cdot n_1) \lim_{x_1\cdot\bar{n}_1\ra\infty} \bigg(\fr{x_1\cdot\bar{n}_1}{n_1\cdot\bar{n}_1}\bigg)^{\De_D} \fr{\<\phi(x_0) \; D(x_1) \; \Ocal(0;n_2)\>}{\lambda_{\phi D\Ocal} \<\Dcal(n_1)\> p^{\tau+\De_\phi-d} (p\cdot n_2)^J}.
\ee
The original position space Wightman function can be written in terms of the building blocks from~\eqref{eq:HandV} as
\be
\<\phi(x_0) \; D(x_1) \; \Ocal(x_2;n_2)\> = \fr{(-1)^{-\half(\De_\phi+\De_D+\De)} \lambda_{\phi D\Ocal} V_{2,01}^J}{x_{01}^{\De_\phi-\De-J+\De_D} x_{12}^{\De+J-\De_\phi+\De_D} x_{02}^{\De_\phi+\De+J-\De_D}}.
\ee
To move the detector to infinity, we first write $x_1$ as the combination of two null vectors,
\be
x_1^\mu = \bigg(\fr{x_1\cdot \bar{n}_1}{n_1\cdot\bar{n}_1}\bigg) n_1^\mu + \bigg(\fr{x_1\cdot n_1}{n_1\cdot\bar{n}_1}\bigg) \bar{n}_1^\mu,
\ee
then take the limit $x_1\cdot\bar{n}_1 \ra \infty$, obtaining
\be
\ba
&\lim_{x_1\cdot\bar{n}_1\ra\infty} \bigg(\fr{x_1\cdot\bar{n}_1}{n_1\cdot\bar{n}_1}\bigg)^{\De_D} \<\phi(x_0) \; D(x_1) \; \Ocal(x_2;n_2) \> \\
& \qquad = \fr{(-1)^{-\half(\De_\phi+\De_D+\De)} \lambda_{\phi D\Ocal} (x_{02}\cdot n_2)^J \Big( 1 - \fr{x_{02}^2(n_1\cdot n_2)}{2(x_{12}\cdot n_1)(x_{02}\cdot n_2)} \Big)^J}{2^{\De_D}(x_{10}\cdot n_1+i\epsilon)^{\half(\De_\phi-\tau+\De_D)} \; (x_{12}\cdot n_1-i\epsilon)^{\half(\tau-\De_\phi+\De_D)} x_{02}^{\De_\phi+\De+J-\De_D}},
\ea
\ee
where the factors of $i\epsilon$ enforce the relative order of the three operators. We can obtain the resulting correlation function of $\Dcal(n_1)$ by expanding the numerator and evaluating integrals of the form
\be
\int_{-\infty}^\infty d(x_1\cdot n_1) \fr{1}{(x_{10}\cdot n_1 + i\epsilon)^a (x_{12}\cdot n_1 - i\epsilon)^b} = (-1)^{a+\half} \fr{2\pi\G(a+b-1)}{\G(a)\G(b)} \fr{1}{(x_{02}\cdot n_1 - i\ep)^{a+b-1}},
\label{eq:BBContour}
\ee
giving us
\be
\ba
\<\phi(x) \; \Dcal(n_1) \; \Ocal(0;n_2) \> &= \fr{(-1)^{\half(J+1)-\De}\lambda_{\phi D\Ocal} \;\pi \G(\De_D-1)}{2^{\De_D-1}\G(\fr{\De_\phi-\tau+\De_D}{2})\;\G(\fr{\tau-\De_\phi+\De_D}{2})} \fr{(x\cdot n_2)^{J}}{(x\cdot n_1)^{\De_D-1} x^{\De_\phi+\De+J-\De_D}} \\
&\qquad \times \sum_{m=0}^{J} (-1)^m \binom{J}{m} \fr{(\De_D-1)_m}{(\fr{\tau-\De_\phi+\De_D}{2})_m} \; y^m,
\ea
\ee
where we've defined the position space cross-ratio
\be
y \equiv \fr{x^2 \; (n_1\cdot n_2)}{2(x\cdot n_1)(x\cdot n_2)}.
\label{eq:CrossRatioPos}
\ee
This sum of terms can be rewritten in closed form as
\be
\ba
&\<\phi(x) \; \Dcal(n_1) \; \Ocal(0;n_2) \> \\
&\qquad = \fr{(-1)^{\half(J+1)-\De}\lambda_{\phi D\Ocal} \;\pi \G(\De_D-1)}{2^{\De_D-1}\G(\fr{\De_\phi-\tau+\De_D}{2})\;\G(\fr{\tau-\De_\phi+\De_D}{2})} \fr{(x\cdot n_2)^{J} \; {}_2F_1(-J,\De_D-1;\fr{\tau-\De_\phi+\De_D}{2};y)}{(x\cdot n_1)^{\De_D-1} x^{\De_\phi+\De+J-\De_D}}.
\ea
\ee
To evaluate the Fourier transform of the source position to momentum space, however, it will be simpler to use the sum representation. Each term in the sum can be Fourier transformed with the general integral~\eqref{eq:BBFourier}, to obtain the momentum space three-point function
\be
\ba
\Ccal_{\phi\Dcal\Ocal}(z) &= \fr{(-1)^{\half(\De_\phi-\tau)} 2^{d+1-\De_\phi-\De} \pi^{\fr{d+2}{2}} \; \G^2(\fr{\De_D}{2}) \; \G(\De_\phi-\fr{\De_D}{2}) \; \G(\De_\phi+\fr{\De_D-d}{2})}{\lambda_{\phi D\phi} \; \G(\De_\phi) \; \G(\De_\phi-\fr{d-2}{2}) \; \G(\fr{\De_\phi-\tau+\De_D}{2})\;\G(\fr{\tau-\De_\phi+\De_D}{2})\;\G(\fr{\De_\phi+\De+J-\De_D}{2})\;\G(\fr{\tau+\De_\phi+\De_D-d}{2})} \\
&\qquad \times \sum_{m=0}^{J} (-1)^m \binom{J}{m} \fr{(\De_D-1)_m \; (\fr{\De_\phi+\De+J-\De_D-2m}{2})_{m}}{(\fr{\tau-\De_\phi+\De_D}{2})_m \; (\fr{\tau+\De_\phi+\De_D-d}{2})_m} \\
&\qquad \quad \times z^m \; {}_2F_1\big(m-J',\De_D-1+m;\tfr{\tau+\De_\phi+\De_D-d+2m}{2};z\big),
\ea
\ee
which can be resummed in terms of a generalized hypergeometric function to the final expression,
\be
\ba
\Ccal_{\phi\Dcal\Ocal}(z) &= \fr{(-1)^{\half(\De_\phi-\tau)} 2^{d+1-\De_\phi-\De} \pi^{\fr{d+2}{2}} \; \G^2(\fr{\De_D}{2}) \; \G(\De_\phi-\fr{\De_D}{2}) \; \G(\De_\phi+\fr{\De_D-d}{2})}{\lambda_{\phi D\phi} \; \G(\De_\phi) \; \G(\De_\phi-\fr{d-2}{2}) \; \G(\fr{\De_\phi-\tau+\De_D}{2})\;\G(\fr{\tau-\De_\phi+\De_D}{2})\;\G(\fr{\De_\phi+\De+J-\De_D}{2})\;\G(\fr{\tau+\De_\phi+\De_D-d}{2})} \\
&\qquad \times {}_3F_2\big(\!-\!J,\De_D-1,\De-1;\tfr{\tau-\De_\phi+\De_D}{2},\tfr{\tau+\De_\phi+\De_D-d}{2};z\big).
\ea
\label{eq:Scalar3ptMom}
\ee

Finally, we need to decompose this three-point function in terms of $SO(d-1)$ representations to obtain the coefficients
\be
\Ccal_{\phi\Dcal\Ocal}^{(\ell)} \equiv \fr{\ell! \; (J-\ell)! \; (\ell+d-3)_{J+1}}{ J! \; 2^J \sqrt{\pi} \; (2\ell+d-3) \; (\fr{d-1}{2})_{J-\half}} \int_0^1 \! \fr{dz}{\big(z(1-z)\big)^{2-\fr{d}{2}}} \; \Chat^{(\fr{d-3}{2})}_\ell(1-2z) \; \Ccal_{\phi\Dcal\Ocal}(z).
\label{eq:PartialWaveIntegral}
\ee
This can be done by expanding~\eqref{eq:Scalar3ptMom} as a polynomial in $z$ and evaluating each integral with~\eqref{eq:BBGegenbauer}, resulting in the sum
\be
\ba
\Ccal_{\phi\Dcal\Ocal}^{(\ell)} &= \fr{(-1)^{\half(\De_\phi-\tau)} \; \pi^{\fr{d+2}{2}} \; \G(d-2+J+\ell) \; \G^2(\fr{\De_D}{2}) \; \G(\De_\phi-\fr{\De_D}{2}) \; \G(\De_\phi+\fr{\De_D-d}{2})}{\lambda_{\phi D\phi} \; 2^{\De_\phi+\De+J-d-1} \G(\De_\phi) \; \G(\De_\phi-\fr{d-2}{2})\;\G(\fr{d-2}{2}+J)\;\G(\fr{\De_\phi-\tau+\De_D}{2})\;\G(\fr{\De_\phi+\De+J-\De_D}{2})} \\
&\qquad \times \sum_{m=\ell}^{J} \binom{J-\ell}{m-\ell} \fr{(-1)^{m-\ell} \; (\De_D-1)_m \; (\De-1)_m \; \G(\fr{d-2}{2}+m)}{\G(\fr{\tau-\De_\phi+\De_D}{2}+m) \; \G(\fr{\tau+\De_\phi+\De_D-d}{2}+m) \; \G(d-2+\ell+m)}.
\ea
\ee
This expression can also be resummed to obtain
\be
\ba
\Ccal_{\phi\Dcal\Ocal}^{(\ell)} &= \fr{(-1)^{\half(\De_\phi-\tau)} \; \pi^{\fr{d+3}{2}} \; \G(d-2+J+\ell) \; \G^2(\fr{\De_D}{2}) \; \G(\De_\phi-\fr{\De_D}{2}) \; \G(\De_\phi+\fr{\De_D-d}{2})}{\lambda_{\phi D\phi} \; 2^{\De_\phi+\De+J+2\ell-4} \G(\De_\phi) \; \G(\fr{2\De_\phi-d+2}{2}) \; \G(\fr{d-2+2J}{2}) \; \G(\fr{d-1+2\ell}{2}) \;\G(\fr{\De_\phi-\tau+\De_D}{2})\;\G(\fr{\De_\phi+\De+J-\De_D}{2})} \\
&\quad \times \fr{(\De-1)_\ell \; \G(\De_D-1+\ell)}{\G(\fr{\tau-\De_\phi+\De_D+2\ell}{2}) \; \G(\fr{\tau+\De_\phi+\De_D-d+2\ell}{2})} \; {}_4 F_3\bigg(\genfrac..{0pt}{}{\ell-J,\De_D-1+\ell,\De-1+\ell,\fr{d-2}{2}+\ell}{\fr{\tau-\De_\phi+\De_D+2\ell}{2},\fr{\tau+\De_\phi+\De_D-d+2\ell}{2},d-2+2\ell}\;;1\bigg).
\ea
\ee
Following~\eqref{eq:ScalarBlock}, we can use these three-point function coefficients to construct the scalar detector conformal block for any intermediate operator $\Ocal$.


\subsection{Energy Operator Matrix Elements}
\label{sec:ThreePtDetails}

The calculation of the energy correlator conformal blocks follows the exact same steps as the scalar detector case presented above. The only complication is that there are multiple tensor structures in the original position space three-point function. Concretely, we have the parity-even structures
\be
\ba
\<\phi(x_0) \; T(x_1;\bar{n}_1) \; \Ocal(x_2;n_2)\>^{(0^+)} &\equiv \fr{(-1)^{-\half(\De_\phi+\De+d)} \lambda^{(0^+)}_{\phi T\Ocal} \; V_{1,02}^2 \; V_{2,01}^{J}}{x_{01}^{\De_\phi-\De-J+d+2} x{12}^{\De+J-\De_\phi+d+2} x_{02}^{\De_\phi+\De+J-d-2}}, \\
\<\phi(x_0) \; T(x_1;\bar{n}_1) \; \Ocal(x_2;n_2)\>^{(1^+)} &\equiv \fr{(-1)^{-\half(\De_\phi+\De+d)} \lambda^{(1^+)}_{\phi T\Ocal} \; V_{1,02} \; V_{2,01}^{J-1} \; H_{12}}{x_{01}^{\De_\phi-\De-J+d+2} x_{12}^{\De+J-\De_\phi+d+2} x_{02}^{\De_\phi+\De+J-d-2}}, \\
\<\phi(x_0) \; T(x_1;\bar{n}_1) \; \Ocal(x_2;n_2)\>^{(2^+)} &\equiv \fr{(-1)^{-\half(\De_\phi+\De+d)} \lambda^{(2^+)}_{\phi T\Ocal} \; V_{2,01}^{J-2} \; H_{12}^2}{x_{01}^{\De_\phi-\De-J+d+2} x_{12}^{\De+J-\De_\phi+d+2} x_{02}^{\De_\phi+\De+J-d-2}},
\ea
\ee
as well as (for $d=3$) the parity-odd structures
\be
\ba
\<\phi(x_0) \; T(x_1;\bar{n}_1) \; \Ocal(x_2;n_2)\>^{(0^-)} &\equiv \fr{(-1)^{-\half(\De_\phi+\De)+1} \lambda^{(0^-)}_{\phi T\Ocal} \; V_{1,02} \; V_{2,01}^{J-1} \; S_{12,0}}{x_{01}^{\De_\phi-\De-J+5} x_{12}^{\De+J-\De_\phi+5} x_{02}^{\De_\phi+\De+J-5}}, \\
\<\phi(x_0) \; T(x_1;\bar{n}_1) \; \Ocal(x_2;n_2)\>^{(1^-)} &\equiv \fr{(-1)^{-\half(\De_\phi+\De)+1} \lambda^{(1^-)}_{\phi T\Ocal} \; V_{2,01}^{J-2} \; H_{12} \; S_{12,0}}{x_{01}^{\De_\phi-\De-J+5} x_{12}^{\De+J-\De_\phi+5} x_{02}^{\De_\phi+\De+J-5}}.
\ea
\ee

Let's consider the parity-even terms first. Taking the limit $x_1\cdot\bar{n}_1 \ra \infty$, these terms become
\be
\ba
&\lim_{x_1\cdot\bar{n}_1\ra\infty} \fr{(x_1\cdot\bar{n}_1)^{d-2}}{(n_1\cdot\bar{n}_1)^d} \<\phi(x_0) \; T(x_1;\bar{n}_1) \; \Ocal(x_2;n_2)\>^{(0^+)} \\
&\qquad = \fr{(-1)^{-\half(\De_\phi+\De+d)} \lambda^{(0^+)}_{\phi T\Ocal} \; (x_{02}\cdot n_1)^2 (x_{02}\cdot n_2)^J \Big( 1 - \fr{x_{02}^2(n_1\cdot n_2)}{2(x_{12}\cdot n_1)(x_{02}\cdot n_2)} \Big)^{J}}{2^{d} \; (x_{10}\cdot n_1 + i\ep)^{\half(\De_\phi-\tau+d+2)} \; (x_{12}\cdot n_1 - i\ep)^{\half(\tau-\De_\phi+d+2)} \; x_{02}^{\De_\phi+\De+J-d+2}}, \\
&\lim_{x_1\cdot\bar{n}_1\ra\infty} \fr{(x_1\cdot\bar{n}_1)^{d-2}}{(n_1\cdot\bar{n}_1)^d} \<\phi(x_0) \; T(x_1;\bar{n}_1) \; \Ocal(x_2;n_2)\>^{(1^+)} \\
&\qquad = \fr{(-1)^{-\half(\De_\phi+\De+d)} \lambda^{(1^+)}_{\phi T\Ocal} \; (n_1\cdot n_2) \; (x_{02}\cdot n_1) \; (x_{02}\cdot n_2)^{J-1} \Big( 1 - \fr{x_{02}^2(n_1\cdot n_2)}{2(x_{12}\cdot n_1)(x_{02}\cdot n_2)} \Big)^{J-1}}{2^{d} \; (x_{10}\cdot n_1 + i\ep)^{\half(\De_\phi-\tau+d)} \; (x_{12}\cdot n_1 - i\ep)^{\half(\tau-\De_\phi+d+4)} \; x_{02}^{\De_\phi+\De+J-d}}, \\
&\lim_{x_1\cdot\bar{n}_1\ra\infty} \fr{(x_1\cdot\bar{n}_1)^{d-2}}{(n_1\cdot\bar{n}_1)^d} \<\phi(x_0) \; T(x_1;\bar{n}_1) \; \Ocal(x_2;n_2)\>^{(2^+)} \\
&\qquad = \fr{(-1)^{-\half(\De_\phi+\De+d)} \lambda^{(2^+)}_{\phi T\Ocal} \; (n_1\cdot n_2)^2 (x_{02}\cdot n_2)^{J-2} \Big( 1 - \fr{x_{02}^2(n_1\cdot n_2)}{2(x_{12}\cdot n_1)(x_{02}\cdot n_2)} \Big)^{J-2}}{2^{d} \; (x_{10}\cdot n_1 + i\ep)^{\half(\De_\phi-\tau+d-2)} \; (x_{12}\cdot n_1 - i\ep)^{\half(\tau-\De_\phi+d+6)} \; x_{02}^{\De_\phi+\De+J-d-2}}.
\ea
\ee
Using~\eqref{eq:BBContour}, we can then integrate each term over $x_1\cdot n_1$ to obtain
\be
\ba
\<\phi(x) \; \Ecal(n_1) \; \Ocal(0;n_2)\>^{(0^+)} \! &= \! \fr{(-1)^{\fr{J-1}{2}-\De} \lambda^{(0^+)}_{\phi T\Ocal} \pi \G(d+1) \; {}_2F_1(J,d+1;\fr{\tau-\De_\phi+d+2}{2};y)}{2^{d-1} \G(\fr{\De_\phi-\tau+d+2}{2}) \G(\fr{\tau-\De_\phi+d+2}{2}) (x\cdot n_1)^{d-1} (x\cdot n_2)^{-J} x^{\De_\phi+\De+J-d+2}}, \\
\<\phi(x) \; \Ecal(n_1) \; \Ocal(0;n_2)\>^{(1^+)} \! &= \! \fr{(-1)^{\fr{J+1}{2}-\De} \lambda^{(1^+)}_{\phi T\Ocal} \pi \G(d+1) (n_1\cdot n_2) \; {}_2F_1(1-J,d+1;\fr{\tau-\De_\phi+d+4}{2};y)}{2^{d-1} \G(\fr{\De_\phi-\tau+d}{2}) \G(\fr{\tau-\De_\phi+d+4}{2}) (x\cdot n_1)^d (x\cdot n_2)^{1-J} x^{\De_\phi+\De+J-d}}, \\
\<\phi(x) \; \Ecal(n_1) \; \Ocal(0;n_2)\>^{(2^+)} \! &= \! \fr{(-1)^{\fr{J-1}{2}-\De} \lambda^{(2^+)}_{\phi T\Ocal} \pi \G(d+1) (n_1\cdot n_2)^2 \; {}_2F_1(2-J,d+1;\fr{\tau-\De_\phi+d+6}{2};y)}{2^{d-1} \G(\fr{\De_\phi-\tau+d-2}{2}) \G(\fr{\tau-\De_\phi+d+6}{2}) (x\cdot n_1)^{d+1} (x\cdot n_2)^{2-J} x^{\De_\phi+\De+J-d-2}},
\ea
\ee
where the cross-ratio $y$ is defined in~\eqref{eq:CrossRatioPos}.

To evaluate the Fourier transform, we again expand each hypergeometric function as a polynomial in $y$ and evaluate each term with~\eqref{eq:BBFourier}, then resum the resulting expressions to find
\be
\ba
\Ccal_{\phi\Ecal\Ocal}^{(0^+)}(z) &= \fr{(-1)^{\half(\De_\phi-\tau)} \; \pi^{d+\fr{3}{2}} \; \G(d+1)}{2^{\De_\phi+\De-2} \G(\fr{d-1}{2}) \; \G(\fr{\De_\phi-\tau+d+2}{2})} \Bigg( \fr{{}_3F_2\big(\!-\!J,d-1,\De+1;\fr{\tau-\De_\phi+d+2}{2},\fr{\tau+\De_\phi+2}{2};z\big)}{\G(\fr{\tau-\De_\phi+d+2}{2}) \; \G(\fr{\tau+\De_\phi+2}{2}) \; \G(\fr{\De_\phi+\De+J-d+2}{2})} \\
&\qquad - \; \fr{2J z \; {}_3F_2\big(1-J,d,\De+1;\fr{\tau-\De_\phi+d+4}{2},\fr{\tau+\De_\phi+4}{2};z\big)}{\G(\fr{\tau-\De_\phi+d+4}{2}) \; \G(\fr{\tau+\De_\phi+4}{2}) \; \G(\fr{\De_\phi+\De+J-d}{2})} \\
&\qquad + \; \fr{J(J-1)z^2 \! {}_3F_2\big(2-J,d+1,\De+1;\fr{\tau-\De_\phi+d+6}{2},\fr{\tau+\De_\phi+6}{2};z\big)}{\G(\fr{\tau-\De_\phi+d+6}{2}) \; \G(\fr{\tau+\De_\phi+6}{2}) \; \G(\fr{\De_\phi+\De+J-d-2}{2})} \Bigg), \\
\Ccal_{\phi\Ecal\Ocal}^{(1^+)}(z) &= \fr{(-1)^{\half(\De_\phi-\tau)+1} \; \pi^{d+\fr{3}{2}} \; \G(d+1)}{2^{\De_\phi+\De-2} \G(\fr{d-1}{2}) \; \G(\fr{\De_\phi-\tau+d}{2})} \Bigg( \fr{z \; {}_3F_2\big(1-J,d,\De+1;\fr{\tau-\De_\phi+d+4}{2},\fr{\tau+\De_\phi+4}{2};z\big)}{\G(\fr{\tau-\De_\phi+d+4}{2}) \; \G(\fr{\tau+\De_\phi+4}{2}) \; \G(\fr{\De_\phi+\De+J-d}{2})} \\
& \qquad - \; \fr{(J-1) z^2 \; {}_3F_2\big(2-J,d+1,\De+1;\fr{\tau-\De_\phi+d+6}{2},\fr{\tau+\De_\phi+6}{2};z\big)}{\G(\fr{\tau-\De_\phi+d+6}{2}) \; \G(\fr{\tau+\De_\phi+6}{2}) \; \G(\fr{\De_\phi+\De+J-d-2}{2})} \Bigg), \\
\Ccal_{\phi\Ecal\Ocal}^{(2^+)}(z) &= \fr{(-1)^{\half(\De_\phi-\tau)} \; \pi^{d+\fr{3}{2}} \; \G(d+1)}{2^{\De_\phi+\De-2} \G(\fr{d-1}{2}) \; \G(\fr{\De_\phi-\tau+d-2}{2})} \; \fr{z^2 \; {}_3F_2\big(2-J,d+1,\De+1;\fr{\tau-\De_\phi+d+6}{2},\fr{\tau+\De_\phi+6}{2};z\big)}{\G(\fr{\tau-\De_\phi+d+6}{2}) \; \G(\fr{\tau+\De_\phi+6}{2}) \; \G(\fr{\De_\phi+\De+J-d-2}{2})}.
\ea
\ee
These three tensor structures can the be combined together, with the OPE coefficients from~\eqref{eq:WardOPE}, to obtain the full parity-even three-point function
\be
\lambda^{(+)}_{\phi\Ecal\Ocal} \; \Ccal_{\phi\Ecal\Ocal}^{(+)}(z) \equiv \lambda^{(0^+)}_{\phi\Ecal\Ocal} \; \Ccal_{\phi\Ecal\Ocal}^{(0^+)}(z) + \lambda^{(1^+)}_{\phi\Ecal\Ocal} \; \Ccal_{\phi\Ecal\Ocal}^{(1^+)}(z) + \lambda^{(2^+)}_{\phi\Ecal\Ocal} \; \Ccal_{\phi\Ecal\Ocal}^{(2^+)}(z),
\ee
resulting in~\eqref{eq:ThreePtMom}.

We can now repeat this procedure for the parity-odd terms. Taking the limit $x_1\cdot\bar{n}_1 \ra \infty$, we have
\be
\ba
&\lim_{x_1\cdot\bar{n}_1\ra\infty} \fr{x_1\cdot\bar{n}_1}{(n_1\cdot\bar{n}_1)^3} \<\phi(x_0) \; T(x_1;\bar{n}_1) \; \Ocal(x_2;n_2)\>^{(0^-)} \\
&\qquad = \fr{(-1)^{-\half(\De_\phi+\De)+1} \lambda^{(0^-)}_{\phi T\Ocal} \; \vep_{\mu\nu\rho} \; x_{02}^\mu \; n_1^\nu \; n_2^\rho \; (x_{02}\cdot n_1) (x_{02} \cdot n_2)^{J-1}\Big(1 - \fr{x_{02}^2 (n_1\cdot n_2)}{2(x_{12}\cdot n_1)(x_{02} \cdot n_2)}\Big)^{J-1}}{8(x_{10}\cdot n_1+i\ep)^{\half(\De_\phi-\tau+4)} \; (x_{12}\cdot n_1-i\ep)^{\half(\tau-\De_\phi+6)} \; x_{02}^{\De_\phi+\De+J-2}}, \\
&\lim_{x_1\cdot\bar{n}_1\ra\infty} \fr{x_1\cdot\bar{n}_1}{(n_1\cdot\bar{n}_1)^3} \<\phi(x_0) \; T(x_1;\bar{n}_1) \; \Ocal(x_2;n_2)\>^{(1^-)} \\
&\qquad = \fr{(-1)^{-\half(\De_\phi+\De)+1} \lambda^{(1^-)}_{\phi T\Ocal} \; \vep_{\mu\nu\rho} \; x_{02}^\mu \; n_1^\nu \; n_2^\rho \; (n_1 \cdot n_2) (x_{02} \cdot n_2)^{J-2} \Big( 1 - \fr{x_{02}^2(n_1\cdot n_2)}{2(x_{12}\cdot n_1)(x_{02} \cdot n_2)}\Big)^{J-2}}{8(x_{10}\cdot n_1+i\ep)^{\half(\De_\phi-\tau+2)} \; (x_{12}\cdot n_1-i\ep)^{\half(\tau-\De_\phi+8)} \; x_{02}^{\De_\phi+\De+J-4}}.
\ea
\ee
Next, we can integrate over $x_1\cdot n_1$ with~\eqref{eq:BBContour}, obtaining
\be
\ba
\<\phi(x) \; \Ecal(n_1) \; \Ocal(0;n_2)\>^{(0^-)} &= \fr{(-1)^{\fr{J-1}{2}-\De} \lambda^{(0^-)}_{\phi T\Ocal} 3\pi \vep_{\mu\nu\rho} x^\mu n_1^\nu n_2^\rho \; {}_2F_1(1-J,4;\fr{\tau-\De_\phi+6}{2};y)}{2\G(\fr{\De_\phi-\tau+4}{2}) \G(\fr{\tau-\De_\phi+6}{2}) (x\cdot n_1)^3 (x\cdot n_2)^{1-J} x^{\De_\phi+\De+J-2}}, \\
\<\phi(x) \; \Ecal(n_1) \; \Ocal(0;n_2)\>^{(1^-)} &= \fr{(-1)^{\fr{J+1}{2}-\De} \lambda^{(1^-)}_{\phi T\Ocal} 3\pi \vep_{\mu\nu\rho} x^\mu n_1^\nu n_2^\rho (n_1 \cdot n_2) \; {}_2F_1(2-J,4;\fr{\tau-\De_\phi+8}{2};y)}{2\G(\fr{\De_\phi-\tau+2}{2}) \G(\fr{\tau-\De_\phi+8}{2})(x\cdot n_1)^4 (x\cdot n_2)^{2-J} x^{\De_\phi+\De+J-4}}.
\ea
\ee
To compute the Fourier transform of these correlators, we need a slight modification of the integral~\eqref{eq:BBFourier},
\be
\ba
&\int d^3x \, e^{i p \cdot x} \fr{\vep_{\mu\nu\rho} \; x^\mu \; n_1^\nu \; n_2^\rho \; (x\cdot n_1)^\alpha (x\cdot n_2)^\beta}{x^{2\de}} \\
&\, = \Theta(p_0) \; \Theta(p^2) \fr{\vep_{\mu\nu\rho} \; p^\mu n_1^\nu n_2^\rho (p\cdot n_1)^{\alpha} (p\cdot n_2)^{\beta} }{p^{2(\alpha+\beta-\de)+5}} \fr{(-1)^{\de-\half(\alpha+\beta+1)}\pi^{\fr{5}{2}} \; {}_2F_1(-\alpha,-\beta;\de-\alpha-\beta-\fr{3}{2};z)}{2^{2\de-\alpha-\beta-5}\G(\de)\G(\de-\alpha-\beta-\fr{3}{2})},
\ea
\ee
with the resulting expressions
\be
\ba
\Ccal_{\phi\Ecal\Ocal}^{(0^-)}(z) &= \fr{(-1)^{\half(\De_\phi-\tau)} \; 3\pi^{\fr{9}{2}}}{2^{\De_\phi+\De-3} \G(\fr{\De_\phi-\tau+4}{2})} \Bigg( \fr{{}_3F_2\big(1-J,3,\De+1;\fr{\tau-\De_\phi+6}{2},\fr{\tau+\De_\phi+3}{2};z\big)}{\G(\fr{\De_\phi-\tau+4}{2}) \; \G(\fr{\tau-\De_\phi+6}{2}) \; \G(\fr{\tau+\De_\phi+3}{2}) \; \G(\fr{\De_\phi+\De+J-2}{2})} \\
&\qquad - \; \fr{(J-1)z \; {}_3F_2\big(2-J,4,\De+1;\fr{\tau-\De_\phi+8}{2},\fr{\tau+\De_\phi+5}{2};z\big)}{\G(\fr{\tau-\De_\phi+8}{2}) \; \G(\fr{\tau+\De_\phi+5}{2}) \; \G(\fr{\De_\phi+\De+J-4}{2})} \Bigg), \\
\Ccal_{\phi\Ecal\Ocal}^{(1^-)}(z) &= \fr{(-1)^{\half(\De_\phi-\tau)+1} \; 3\pi^{\fr{9}{2}}}{2^{\De_\phi+\De-3} \G(\fr{\De_\phi-\tau+2}{2})} \; \fr{z \; {}_3F_2\big(2-J,4,\De+1;\fr{\tau-\De_\phi+8}{2},\fr{\tau+\De_\phi+5}{2};z\big)}{\G(\fr{\tau-\De_\phi+8}{2}) \; \G(\fr{\De_\phi+\De+J-4}{2}) \; \G(\fr{\tau+\De_\phi+5}{2})}.
\ea
\ee
If we combine these two terms together, with the OPE coefficients from~\eqref{eq:WardOPEOdd}, we obtain the full parity-odd three-point function
\be
\lambda^{(-)}_{\phi\Ecal\Ocal} \; \Ccal_{\phi\Ecal\Ocal}^{(-)}(z) \equiv \lambda^{(0^-)}_{\phi\Ecal\Ocal} \; \Ccal_{\phi\Ecal\Ocal}^{(0^-)}(z) + \lambda^{(1^-)}_{\phi\Ecal\Ocal} \; \Ccal_{\phi\Ecal\Ocal}^{(1^-)}(z),
\ee
shown in~\eqref{eq:ThreePtMomOdd}.


\subsection{Decomposition into Partial Waves}
\label{sec:CoeffDetails}

In order to use these three-point functions to construct conformal blocks, we need to decompose them into $SO(d-1)$ representations. The procedure for the parity-even tensor structures is the same as for the scalar detector example, where we simply need to integrate~\eqref{eq:ThreePtMom} against Gegenbauer polynomials, as in~\eqref{eq:PartialWaveIntegral}.

Because not all tensor structures are allowed for intermediate operator $\Ocal$ with spin $J<2$, we need to consider those cases separately. Starting with $J=0$, we have
\be
\Ccal^{(+)}_{\phi\Ecal\Ocal}(z) = \Ccal^{(0^+)}_{\phi\Ecal\Ocal}(z) = \fr{(-1)^{\half(\De_\phi-\De)} \; \pi^{d+\fr{3}{2}} \; \G(d+1)}{2^{\De_\phi+\De-2} \; \G(\fr{d-1}{2}) \; \G(\fr{\De_\phi-\De+d+2}{2}) \; \G(\fr{\De-\De_\phi+d+2}{2}) \; \G(\fr{\De+\De_\phi+2}{2}) \; \G(\fr{\De_\phi+\De-d+2}{2})}.
\ee
Since $\Pi^0_0(z) = 1$, this correlator is the same as the $\ell=0$ coefficient given in~\eqref{eq:ThreePtL0}.

Next we have the case $J=1$,
\be
\ba
\Ccal^{(+)}_{\phi\Ecal\Ocal}(z) &= \Ccal_{\phi\Ecal\Ocal}^{(0^+)}(z) - \fr{2\big( (d-1)(\De-\De_\phi)+1\big)}{(d-2)(\De-\De_\phi-d-1)} \; \Ccal_{\phi\Ecal\Ocal}^{(1^+)}(z) \\
&= \fr{(-1)^{\half(\De_\phi-\tau)} \; \pi^{d+\fr{3}{2}} \; \G(d+1)}{2^{\De_\phi+\De-2} \; \G(\fr{d-1}{2}) \; \G(\fr{\De_\phi-\De+d+3}{2}) \; \G(\fr{\De_\phi+\De-d+3}{2})} \Bigg( \fr{1}{\G(\fr{\De-\De_\phi+d+1}{2}) \; \G(\fr{\De+\De_\phi+1}{2})} \\
&\qquad - \; \fr{(d-1)\big(\De_\phi(d-\De_\phi)+\De^2-1\big)z}{(d-2)\G(\fr{\De-\De_\phi+d+3}{2}) \; \G(\fr{\De+\De_\phi+3}{2})} \Bigg).
\ea
\ee
Decomposing this expression in terms of $\Pi^1_0(z) = 1$ and $\Pi^1_1(z) = 1-2z$, we obtain the $\ell=0$ coefficient in~\eqref{eq:ThreePtL0} and the $\ell=1$ coefficient in~\eqref{eq:ThreePtL1}.

The remaining parity-even partial wave coefficients for $J\geq2$ can be obtained by expanding~\eqref{eq:ThreePtMom} in powers of $z$, integrating each term against the appropriate Gegenbauer polynomial using~\eqref{eq:BBGegenbauer}, then resumming into a combination of hypergeometric functions, resulting in the final expression~\eqref{eq:ThreePtL}.

For the parity-odd partial wave coefficients allowed in $d=3$, we can simplify the calculation by going to the rest frame of the external state, such that $z = \half(1-\cos\theta)$. In these coordinates, we can rewrite the three-point functions~\eqref{eq:ThreePtMomOdd} in the form
\be
\fr{p \; \vep_{\mu\nu\rho} \; p^\mu n_1^\nu n_2^\rho}{(p\cdot n_1) (p\cdot n_2)} \; \Ccal^{(-)}_{\phi\Ecal\Ocal}(z) = \sin\theta \, \Ccal^{(-)}_{\phi\Ecal\Ocal}(z) \equiv \Ccal^{(-)}_{\phi\Ecal\Ocal}(\theta).
\ee
We now need to decompose this function in terms of $SO(2)$ representations. Instead of the usual parity-even basis, which in $d=3$ takes the form
\be
\Pi_{\ell^+}^J(\theta) = \fr{(2J)!}{2^{J-1}\;(J-\ell)!\;(J+\ell)!} \; \cos\ell\theta,
\ee
we need to use parity-odd basis,
\be
\Pi_{\ell^-}^J(\theta) = \fr{(2J)!}{2^{J-1}\;(J-\ell)!\;(J+\ell)!} \; \sin\ell\theta,
\ee
to obtain the expansion
\be
\Ccal^{(-)}_{\phi\Ecal\Ocal}(\theta) = \sum_{\ell=1}^J \Ccal^{(\ell^-)}_{\phi\Ecal\Ocal} \; \Pi_{\ell^-}^J(\theta).
\ee

Because only one parity-odd tensor structure is allowed for $J=1$, we need to treat that case separately, with the associated three-point function
\be
\ba
\Ccal^{(-)}_{\phi\Ecal\Ocal}(\theta) &= \fr{(-1)^{\half(\De_\phi-\tau)} \; 3\pi^{\fr{9}{2}} \; \sin\theta}{2^{\De_\phi+\De-3} \; \G(\fr{\De_\phi-\De+5}{2}) \; \G(\fr{\De-\De_\phi+5}{2}) \; \G(\fr{\De_\phi+\De-1}{2}) \; \G(\fr{\De+\De_\phi+2}{2})},
\ea
\ee
allowing us to easily read off the $\ell=1$ coefficient~\eqref{eq:ThreePtL1Odd}.

The remaining parity-odd partial wave coefficients can be computed by integrating~\eqref{eq:ThreePtMomOdd} against the $SO(2)$ basis,
\be
\Ccal^{(\ell^-)}_{\phi\Ecal\Ocal} = \fr{2^{J-1} (J-\ell)! \; (J+\ell)!}{(2J)! \; \pi} \int_0^{2\pi} \!\!d\theta \; \sin\ell\theta \, \Ccal^{(-)}_{\phi\Ecal\Ocal}(\theta),
\ee
resulting in~\eqref{eq:ThreePtLOdd}.


\section{OPE Coefficients from Conglomeration}
\label{app:Conglomeration}

In this appendix we use the method of conglomeration developed in~\cite{Fitzpatrick:2011dm} to compute the OPE coefficients for the tensor product CFT example of Sec.~\ref{sec:ProductCFT}. Note that while the overall focus of this paper is the computation of energy correlators, which necessarily requires Lorentzian signature, here we just want to extract OPE coefficients, so to simplify the calculation we will work in Euclidean signature.


\subsection{Overview of Conglomeration Method}

Our goal is to compute OPE coefficients involving the double-twist operators $[\phi_1\phi_2]_{n,J}$ built from two scalar primary operators with the same scaling dimension $\De_\phi$. While these operators are typically considered in the context of generalized free theories, we are instead studying a tensor product theory built from two copies of the same CFT.

The basic idea of conglomeration (see~\cite{Fitzpatrick:2011dm} for details) is that these double-twist operators can be built from the constituent operators via the integral
\be
[\phi_1\phi_2]_{n,J}(x_0;n_0) = \int d^dx_1 \; d^dx_2 \; f_{n,J}(x_0;n_0|x_1,x_2) \; \phi_1(x_1) \; \phi_2(x_2),
\label{eq:Conglomeration}
\ee
where the ``wavefunction'' $f_{n,J}$ is given by\footnote{While in this appendix we are working in Euclidean signature, we analytically continue all auxiliary vectors $n_i$ to complex values such that $n_i^2 = 0$.}
\be
f_{n,J}(x_0;n_0|x_1,x_2) \equiv \fr{1}{N_{n,J}} \; \fr{\big(x_{01}^2 (x_{02}\cdot n_0) - x_{02}^2 (x_{01}\cdot n_0) \big)^J}{x_{01}^{2\De_\phi+2n+2J} \; x_{02}^{2\De_\phi+2n+2J} x_{12}^{2d-4\De_\phi-2n}}.
\ee
The normalization coefficient $N_{n,J}$, which ensures that the resulting double-twist operator's two-point function is of the form~\eqref{eq:TwoPtPos}, is given by
\be
\ba
N_{n,J}^2 &= \fr{\pi^{2d} \; n! \; J! \; \G(2\De_\phi+n+J-\fr{d}{2}) \; \G(2\De_\phi+2n+J-\fr{d}{2}) \; (2\De_\phi+n-d+1)_n}{2^J \; \G^2(\De_\phi) \; \G^2(\De_\phi+n+J) \; \G(J+\fr{d}{2}) \; \G(n+J+\fr{d}{2})} \\
&\qquad \times \fr{\G^2(-n) \; \G^2(\fr{d}{2}-\De_\phi) \; \G^2(\fr{d}{2}-\De_\phi-n) \; (2\De_\phi+2n+J-1)_J}{\G^2(d-2\De_\phi-n)}. \fr{}{}
\ea
\ee
Attentive readers may notice that this coefficient is formally infinite, due to the factor of $\G(-n)$. As discussed in~\cite{Fitzpatrick:2011dm}, this divergence can be formally regulated by analytically continuing the dimension of the double-trace operator to $\De_{n,J} = 2\De_\phi+2n+J+\epsilon$ until the end of the calculation, where the resulting OPE coefficients are finite in the limit $\ep\ra0$. For notational simplicity, we will ignore these factors of $\ep$ and only restrict to $n\in\mathbb{N}$ in the final result.

\begin{figure}[t!]
\centering
\includegraphics[width=.8\linewidth]{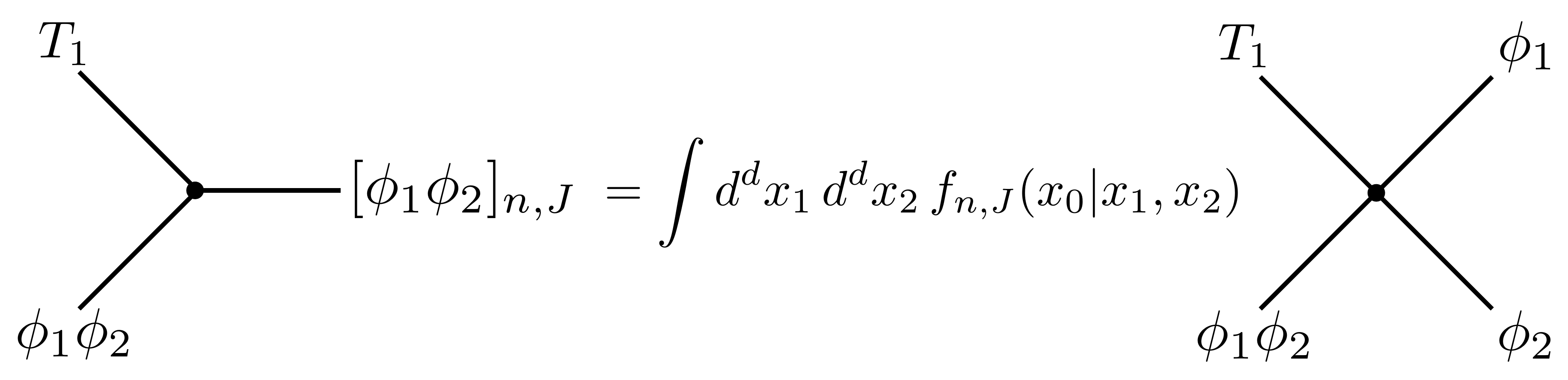}
\caption{Three-point functions involving the double-twist operator $[\phi_1\phi_2]_{n,J}(x_0)$ can be obtained from a four-point function containing $\phi_1(x_1)$ and $\phi_2(x_2)$ by ``conglomerating'' the two constituent operators with~\eqref{eq:Conglomeration}.}
\label{fig:Conglomeration}
\end{figure}

To compute double-twist OPE coefficients, we can apply~\eqref{eq:Conglomeration} to a four-point function containing $\phi_1$ and $\phi_2$ to obtain a three-point function with $[\phi_1\phi_2]_{n,J}$, as shown schematically in Figure~\ref{fig:Conglomeration},
\be
\ba
&\<\phi_1\phi_2(x_0) \; T_1(x_1;n_1) \; [\phi_1\phi_2]_{n,J}(x_2;n_2)\> \\
& \qquad = \int d^dx_3 \; d^dx_4 \; f_{n,J}(x_2;n_2|x_3,x_4) \; \<\phi_1\phi_2(x_0) \; T_1(x_1;n_1) \; \phi_1(x_3) \; \phi_2(x_4)\>,
\ea
\ee
with an analogous expression for correlators involving $T_2$.\footnote{Using the relation $f_{n,J}(x_0;n_0|x_2,x_1) = (-1)^J f_{n,J}(x_0;n_0|x_1,x_2)$, one can show that
\be
\lambda_{(\phi_1\phi_2)T_2[\phi_1\phi_2]_{n,J}} = (-1)^J \lambda_{(\phi_1\phi_2)T_1[\phi_1\phi_2]_{n,J}}.
\ee
}
Because the two copies of the underlying CFT are not coupled to each other, we can easily compute the required four-point function from a product of lower-point functions,
\be
\<\phi_1\phi_2(x_0) \; T_1(x_1;n_1) \; \phi_1(x_3) \; \phi_2(x_4)\> = \<\phi_1(x_0) \; T_1(x_1;n_1) \; \phi_1(x_3)\> \; \<\phi_2(x_0) \; \phi_2(x_4)\>.
\ee

However, we still need to evaluate the integrals over $x_3$ and $x_4$. This can be done with a generalization of the Symanzik ``star formula''~\cite{Symanzik:1972wj} derived in~\cite{Chen:2017xdz} (see also~\cite{Paulos:2012nu,Giombi:2017hpr}),
\be
\int d^dx_0 \; \prod_{i=1}^k \fr{\G(\ell_i)}{\xi_i!} \fr{(x_{i0} \cdot n_i)^{\xi_i}}{x_{i0}^{2\ell_i}} = \pi^{\fr{d}{2}} \sum_{\{a,b\}} \int [d\de] \prod_{i\neq j} \fr{(x_{ij} \cdot n_i)^{a_{ij}}}{a_{ij}!} \prod_{i<j} \fr{\big(\half \; n_i \cdot n_j\big)^{b_{ij}}}{b_{ij}!} \fr{\G(\de_{ij})}{x_{ij}^{2\de_{ij}}},
\label{eq:GenStarFormula}
\ee
which holds for all $\ell_i\in\mathbb{R}$ and $\xi_i\in\mathbb{N}$ satisfying
\be
\sum_{i=1}^k (\ell_i - \xi_i) = d,
\ee
so long as all $n_i^2 = 0$. The general formula~\eqref{eq:GenStarFormula} requires a bit of unpacking. The summation variables $a_{ij}$ and $b_{ij}$ run over nonnegative integers,
\be
a_{ij} \geq 0, \quad b_{ij} \geq 0,
\ee
with upper bounds set by the constraints
\be
\sum_{j\neq i} (a_{ij} + b_{ij}) = \xi_i.
\ee
Both $b_{ij}$ and the integration variables $\de_{ij}$ are symmetric ($b_{ji} = b_{ij}, \; \de_{ji} = \de_{ij}$), while the $a_{ij}$ are not (generically $a_{ji} \neq a_{ij}$), and the diagonal elements of all three variables are zero ($a_{ii} = b_{ii} = \de_{ii} = 0$). Finally, the integration measure for the $\de_{ij}$ is evaluated over imaginary values,
\be
\int[d\de] \equiv \prod_{[ij]} \int_{-i\infty}^{i\infty} \fr{d\de_{ij}}{2\pi i},
\ee
where $[ij]$ runs over the $\half k(k-3)$ independent variables satisfying the constraints
\be
\sum_{j\neq i} (\de_{ij} - a_{ji}) = \ell_i.
\ee
Fortunately, we will only need the case $k=3$, where these constraints completely fix all $\de_{ij}$, with no remaining integrals to evaluate. By construction, we end up with a three-point function in the form~\eqref{eq:ThreePtPos} required by conformal invariance, from which we can read off the desired OPE coefficient.


\subsection{Warmup: Scalar Detector}

To gain some intuition for the conglomeration procedure, we'll first consider the simpler example of OPE coefficients for a scalar primary operator $D$ with scaling dimension $\De_D$, before moving on to our eventual goal of the stress tensor $T_{\mu\nu}$. Concretely, we'd like to compute the Euclidean three-point function
\be
\ba
&\<\phi_1\phi_2(x_0) \; D_1(x_1) \; [\phi_1\phi_2]_{n,J}(x_2;n_2)\> \equiv \fr{\lambda_{n,J} \big(x_{12}^2 (x_{02} \cdot n_2) - x_{02}^2(x_{12} \cdot n_2)\big)^J}{x_{01}^{\De_D-2n} x_{12}^{\De_D+2n+2J} x_{02}^{4\De_\phi+2n+2J-\De_D}} \\
&\hspace{3.3cm} = \int d^dx_3 \; d^dx_4 \; f_{n,J}(x_2;n_2|x_3,x_4) \; \<\phi_1\phi_2(x_0) \; D_1(x_1) \; \phi_1(x_3) \; \phi_2(x_4)\>,
\ea
\ee
where we've used the shorthand notation $\lambda_{n,J} \equiv \lambda_{(\phi_1\phi_2)D_1[\phi_1\phi_2]_{n,J}}$. As we're only interested in the overall OPE coefficient, we can simplify the resulting calculations by choosing the operator positions $x_i$ such that $x_{02}\cdot n_2 = 0$.

The four-point function in the integrand is given by
\be
\ba
\<\phi_1\phi_2(x_0) \; D_1(x_1) \; \phi_1(x_3) \; \phi_2(x_4)\> &= \<\phi_1(x_0) \; D_1(x_1) \; \phi_1(x_3)\> \; \<\phi_2(x_0) \; \phi_2(x_4)\> \\
&= \fr{\lambda_{\phi D\phi}}{x_{01}^{\De_D} \; x_{03}^{2\De_\phi-\De_D} \; x_{13}^{\De_D} \; x_{04}^{2\De_\phi}},
\ea
\ee
so the double-twist OPE coefficient is obtained from the integral
\be
\lambda_{n,J} = \fr{(-1)^J \lambda_{\phi D\phi} \; x_{12}^{\De_D+2n+2J}}{N_{n,J}(x_{12} \cdot n_2)^J x_{01}^{2n} \; x_{02}^{\De_D-4\De_\phi-2n}} \int d^dx_3 \; d^dx_4 \; \fr{x_{03}^{\De_D-2\De_\phi} \big(x_{23}^2 (x_{24}\cdot n_2) - x_{24}^2 (x_{23}\cdot n_2) \big)^J}{x_{13}^{\De_D} x_{23}^{2\De_\phi+2n+2J} x_{04}^{2\De_\phi} x_{24}^{2\De_\phi+2n+2J} x_{34}^{2d-4\De_\phi-2n}}.
\ee
We can expand the numerator to rewrite this as a sum of integrals
\be
\ba
&\int d^dx_3 \; d^dx_4 \; \fr{x_{03}^{\De_D-2\De_\phi} \big(x_{23}^2 (x_{24}\cdot n_2) - x_{24}^2 (x_{23}\cdot n_2) \big)^J}{x_{13}^{\De_D} x_{23}^{2\De_\phi+2n+2J} x_{04}^{2\De_\phi} x_{24}^{2\De_\phi+2n+2J} x_{34}^{2d-4\De_\phi-2n}} \\
&\qquad = \sum_{m=0}^J (-1)^m \binom{J}{m} \int d^dx_3 \; \fr{x_{03}^{\De_D-2\De_\phi} (x_{23}\cdot n_2)^m}{x_{13}^{\De_D} x_{23}^{2\De_\phi+2n+2m}} \int d^dx_4 \; \fr{x_{34}^{4\De_\phi+2n-2d} (x_{24}\cdot n_2)^{J-m}}{x_{04}^{2\De_\phi} x_{24}^{2\De_\phi+2n+2(J-m)}}.
\ea
\ee
The individual integrals can be evaluated with the generalized star formula~\eqref{eq:GenStarFormula}, starting with the one over $x_4$,
\be
\ba
&\int d^dx_4 \; \fr{x_{34}^{4\De_\phi+2n-2d} (x_{24}\cdot n_2)^{J-m}}{x_{04}^{2\De_\phi} x_{24}^{2\De_\phi+2n+2(J-m)}} \\
&\qquad = \fr{\pi^{\fr{d}{2}} \; \G(2\De_\phi+n-\fr{d}{2}) \; \G(\fr{d}{2}-\De_\phi-n) \; \G(\fr{d}{2}-\De_\phi+J-m)}{\G(\De_\phi) \; \G(\De_\phi+n+J-m) \; \G(d-2\De_\phi-n)} \; \fr{x_{03}^{2\De_\phi+2n-d} (x_{23} \cdot n_2)^{J-m}}{x_{02}^{4\De_\phi+2n-d} x_{23}^{d-2\De_\phi+2(J-m)}},
\ea
\label{eq:ConglomerationX4}
\ee
which has been significantly simplified by our choice to set $x_{02} \cdot n_2 = 0$. We can again use~\eqref{eq:GenStarFormula} to evaluate the resulting integral over $x_3$,
\be
\int d^dx_3 \; \fr{x_{03}^{\De_D+2n-d} (x_{23}\cdot n_2)^J}{x_{13}^{\De_D} x_{23}^{2n+2J+d}} = \fr{(-1)^J \pi^{\fr{d}{2}} \; \G(-n) \; \G(n+J+\fr{\De_D}{2}) \; \G(\fr{d-\De_D}{2})}{\G(\fr{\De_D}{2}) \; \G(n+J+\fr{d}{2}) \; \G(\fr{d-\De_D}{2}-n)} \; \fr{x_{01}^{2n} (x_{12} \cdot n_2)^J}{x_{02}^{d-\De_D} x_{12}^{\De_D+2n+2J}}.
\ee

Combining these expressions together, we can write the resulting OPE coefficient as the sum
\be
\ba
\lambda_{n,J} &= \fr{\lambda_{\phi D\phi}}{N_{n,J}} \; \fr{(-1)^n \pi^d \; \G(-n) \; \G(2\De_\phi+n-\fr{d}{2}) \; (\fr{\De_D-d+2}{2})_n \; (\fr{\De_D}{2})_{n+J} \; \G(\fr{d}{2}-\De_\phi-n)}{\G(\De_\phi) \; \G(n+J+\fr{d}{2}) \; \G(d-2\De_\phi-n)} \\
&\qquad \times \sum_{m=0}^J (-1)^m \binom{J}{m} \fr{\G(\fr{d}{2}-\De_\phi+J-m)}{\G(\De_\phi+n+J-m)}.
\ea
\ee
Note the factor of $\G(-n)$ in the numerator, which cancels against the equivalent factor in $N_{n,J}$. We can resum this result to obtain the finite OPE coefficients
\be
\lambda^2_{n,J} = \fr{\lambda^2_{\phi D\phi} \; 2^J (\fr{\De_D-d+2}{2})_n^2 \; (\fr{\De_D}{2})_{n+J}^2}{n! \; J! \; (J+\fr{d}{2})_n \; (2\De_\phi+n-d+1)_n \; (2\De_\phi+n+J-\fr{d}{2})_n \; (2\De_\phi+2n+J-1)_J}.
\ee
As a sanity check, for $\De_D = 2\De_\phi$ we recover the well-known generalized free theory OPE coefficients derived in~\cite{Fitzpatrick:2011dm} (multiplied by $\lambda_{\phi D\phi}$).\footnote{This specific case (i.e.~$D=\phi^2$) was previously considered in~\cite{Kologlu:2019bco}.}


\subsection{Stress Tensor OPE Coefficients}

We can now use the same procedure to compute three-point functions containing the stress tensor $T_{\mu\nu}$,
\be
\ba
&\<\phi_1\phi_2(x_0) \; T_1(x_1;n_1) \; [\phi_1\phi_2]_{n,J}(x_2;n_2)\> \\
&\quad \equiv \fr{\lambda_{n,J} \big( V_{1,02}^2 \; V_{2,01}^J -2\big(\fr{(d-1)(2n+J)+J}{(d-2)(2n-d)}\big) V_{1,02} \; V_{2,01}^{J-1} \; H_{12} + \big(\fr{(d-1)(2n+J)^2-(d-2)J-J^2}{(d-2)(2n-d)(2n-d+2)}\big) V_{2,01}^{J-2} \; H_{12}^2 \big)}{x_{01}^{d+2-2n-2J} x_{12}^{d+2+2n+2J} x_{02}^{4\De_\phi+2n+2J-d-2}},
\ea
\label{eq:DoubleTwist3pt}
\ee
where we've again used the shorthand $\lambda_{n,J} \equiv \lambda_{(\phi_1\phi_2)T_1[\phi_1\phi_2]_{n,J}}$. To simplify the calculations, we will position the operators such that $x_{01}\cdot n_1 = x_{02}\cdot n_2 = 0$.

Just like for the scalar case, we need to first compute the four-point function
\be
\<\phi_1\phi_2(x_0) \; T_1(x_1;n_1) \; \phi_1(x_3) \; \phi_2(x_4)\> = \fr{\De_\phi \; \G(\fr{d+2}{2})}{\pi^{\fr{d}{2}}(d-1)} \; \fr{(x_{13}\cdot n_1)^2}{x_{01}^{d-2} \; x_{03}^{2\De_\phi-d+2} \; x_{13}^{d+2} \; x_{04}^{2\De_\phi}}.
\ee
We can then obtain the double-twist three-point function, and the associated OPE coefficient, from the integral
\be
\ba
&\<\phi_1\phi_2(x_0) \; T_1(x_1;n_1) \; [\phi_1\phi_2]_{n,J}(x_2;n_2)\> \\
&\quad = \fr{\De \; \G(\fr{d+2}{2})}{\pi^{\fr{d}{2}}(d-1) \; N_{n,J}} \; \fr{1}{x_{01}^{d-2}} \int d^dx_3 \; d^dx_4 \; \fr{x_{03}^{d-2-2\De_\phi} (x_{13}\cdot n_1)^2 \big(x_{23}^2 (x_{24}\cdot n_2) - x_{24}^2 (x_{23}\cdot n_2) \big)^{J}}{x_{13}^{d+2} x_{23}^{2\De_\phi+2n+2J} x_{04}^{2\De_\phi} x_{24}^{2\De_\phi+2n+2J} x_{34}^{2d-4\De_\phi-2n}},
\ea
\ee
which we can again rewrite as the sum
\be
\ba
&\int d^dx_3 \; d^dx_4 \; \fr{x_{03}^{d-2-2\De_\phi} (x_{13}\cdot n_1)^2 \big(x_{23}^2 (x_{24}\cdot n_2) - x_{24}^2 (x_{23}\cdot n_2) \big)^J}{x_{13}^{d+2} x_{23}^{2\De_\phi+2n+2J} x_{04}^{2\De_\phi} x_{24}^{2\De_\phi+2n+2J} x_{34}^{2d-4\De_\phi-2n}} \\
&\quad = \sum_{m=0}^J (-1)^m \binom{J}{m} \int d^dx_3 \; \fr{x_{03}^{d-2-2\De_\phi} (x_{13}\cdot n_1)^2 (x_{23}\cdot n_2)^m}{x_{13}^{d+2} x_{23}^{2\De_\phi+2n+2m}} \int d^dx_4 \; \fr{x_{34}^{4\De_\phi+2n-2d} (x_{24}\cdot n_2)^{J-m}}{x_{04}^{2\De_\phi} x_{24}^{2\De_\phi+2n+2(J-m)}}.
\ea
\ee
The integral over $x_4$ is the same as the scalar case in~\eqref{eq:ConglomerationX4}, so we only need to evaluate the resulting integral over $x_3$ with~\eqref{eq:GenStarFormula},
\be
\ba
&\int d^dx_3 \; \fr{x_{03}^{2n-2} (x_{13}\cdot n_1)^2 (x_{23}\cdot n_2)^J}{x_{13}^{d+2} x_{23}^{2n+2J+d}} \\
&\qquad =\fr{(-1)^J \; \pi^{\fr{d}{2}} \; \G(-n) \big( (n+J+\fr{d}{2})(n+J+\fr{d-2}{2}) - 2J(n+J+\fr{d-2}{2}) \; y_{12} + J(J-1) \; y_{12}^2 \big)}{(n+J+\fr{d-2}{2}) \; \G(1-n) \; \G(\fr{d+2}{2}) \; (x_{12}\cdot n_1)^{-2} (x_{12}\cdot n_2)^{-J} x_{01}^{-2n} \; x_{02}^2 \; x_{12}^{2n+2J+d+2}},
\ea
\ee
where $y_{12}$ is the cross-ratio defined in~\eqref{eq:CrossRatioPos} with $x=x_{12}$.

We can combine all these pieces together to reconstruct the full three-point function
\be
\ba
&\<\phi_1\phi_2(x_0) \; T_1(x_1;n_1) \; [\phi_1\phi_2]_{n,J}(x_2;n_2)\>  \\
&\qquad= \fr{(-1)^J \; \pi^{\fr{d}{2}} (n+\fr{d}{2})(n+\fr{d-2}{2}) \; \De_\phi \; \G(-n) \; \G(2\De_\phi+n+J-\fr{d}{2}) \; \G(\fr{d}{2}-\De_\phi) \; \G(\fr{d}{2}-\De_\phi-n)}{N_{n,J} \; (d-1) (n+J+\fr{d-2}{2}) \; \G(1-n) \; \G(\De_\phi) \; \G(\De_\phi+n+J) \; \G(d-2\De_\phi-n)} \\
&\qquad \quad \times \fr{\big( V_{1,02}^2 \; V_{2,01}^J + \fr{2J}{(2n+d-2)} V_{1,02} \; V_{2,01}^{J-1} H_{12} + \fr{J(J-1)}{(2n+d)(2n+d-2)} V_{2,01}^{J-2} H_{12}^2 \big)}{x_{01}^{d+2-2n-2J} \; x_{12}^{d+2+2n+2J} \; x_{02}^{4\De+2n+2J-d-2}},
\ea
\label{eq:Conglomeration3pt}
\ee
with the OPE coefficient
\be
\ba
\lambda^2_{n,J} &= \fr{2^J \; \De_\phi^2 (n+\fr{d}{2})^2 (n+\fr{d-2}{2})^2 \; \G(J+\fr{d}{2}) \; \G(n+J+\fr{d-2}{2})}{n! \; J! \; (2\De_\phi+n-d+1)_n \; (2\De_\phi+n+J-\fr{d}{2})_n \; (2\De_\phi+2n+J-1)_J} \\
&\qquad \times \fr{1}{\pi^d (d-1)^2 (n+J+\fr{d-2}{2}) \; \G^2(1-n)}.
\ea
\ee
Unlike the scalar case, we find a factor of $\G(1-n)$ in the denominator, which kills all OPE coefficients with $n\neq 0$, leaving us with the much simpler expression presented in~\eqref{eq:ProductCFTOPE}. This vanishing of OPE coefficients appears to be required by the Ward identity, as the three-point function~\eqref{eq:Conglomeration3pt} we obtained with conglomeration has the \emph{wrong} relative coefficients between the three tensor structures, in comparison with~\eqref{eq:DoubleTwist3pt}, for all $n \neq 0$.


\bibliographystyle{utphys}
\bibliography{References}

\end{document}